\documentclass{aa}  

\usepackage{graphicx}
\usepackage{txfonts}
\usepackage[margin=0.2cm]{caption}

\usepackage{color}

\begin{document}

   \title{Type~II Cepheids in the Milky Way disc\thanks{Based on observations collected at the European Organisation for Astronomical Research in the Southern Hemisphere, Chile (prog. ID: 060.A-9120 and 082.D-0901)}} 

   \subtitle{Chemical composition of two new W~Vir stars: DD~Vel and HQ~Car}

   \author{B. Lemasle
          \inst{1}
          \and
          V. Kovtyukh\inst{2}
          \and
          G. Bono\inst{3}
          \and
          P. Fran\c cois\inst{4,5}
          \and
          I. Saviane\inst{6}
          \and
          I. Yegorova\inst{6}
          \and
          K. Genovali\inst{3}
          \and
          L. Inno\inst{3,7} 
          \and          \newline
          G. Galazutdinov\inst{8,9}
          \and
          R. da Silva\inst{3}
}
          \institute{Anton Pannekoek Institute for Astronomy, University of Amsterdam, Science Park 904, PO Box 94249, 1090 GE, Amsterdam, The Netherlands, 
              \email{B.J.P.Lemasle@uva.nl}
          \and
          Astronomical Observatory, Odessa National University, and Isaac Newton Institute of Chile, Odessa branch, Shevchenko Park, 65014, Odessa, Ukraine
          \and
          Dipartimento di Fisica, Universit\`a di Roma Tor Vergata, via della Ricerca Scientifica 1, 00133 Rome, Italy
          \and
          GEPI, Observatoire de Paris, CNRS, Universit\'e Paris Diderot, Place Jules Janssen, 92190 Meudon, France
          \and
          UPJV-Universit\'e de Picardie Jules Verne, 80000 Amiens, France
          \and
          European Southern Observatory, Alonso de Cordova 3107, Santiago, Chile
          \and
          European Southern Observatory, Karl-Schwarzschild-Str. 2, D-85748 Garching bei Munchen, Germany
          \and
          Instituto de Astronomia, Universidad Catolica del Norte, av. Angamos 0610, Antofagasta, Chile
          \and
          Pulkovo Observatory, Pulkovskoe Shosse 65, Saint-Petersburg 196140, Russia
}

   \date{Received September 15, 1996; accepted March 16, 1997}

  \abstract
   {A robust classification of Cepheids into their different sub-classes and, in particular, between classical and Type II Cepheids, is necessary to properly calibrate the period-luminosity relations and for populations studies in the Galactic disc. Type II Cepheids are, however, very diverse, and classifications based either on intrinsic (period, light curve) or external parameters (e.g., [Fe/H], $|z|$) do not provide a unique classification.}
   {We want to ascertain the classification of two Cepheids, HQ~Car and DD~Vel, that are sometimes classified as classical Cepheids and sometimes as Type~II Cepheids.}
   {To achieve this goal, we examine both their chemical composition and the presence of specific features in their spectra.}
   {We find emission features in the H$\alpha$ and in the 5875.64 \AA{} He~I lines that are typical of W~Vir stars. The [Na/Fe] (or [Na/Zn]) abundances are typical of thick-disc stars, while BL~Her stars are Na-overabundant ([Na/Fe]$>$+0.5 dex). Finally, the two Cepheids show a possible (HQ~Car) or probable (DD~Vel) signature of mild dust-gas separation that is usually observed only in long-period type~II Cepheids and RV~Tau stars.}
   {These findings clearly indicate that HQ~Car and DD~Vel are both Type~II Cepheids from the W~Vir sub-class. Several studies have reported an increase in the Cepheids' abundance dispersion towards the outer (thin) disc. A detailed inspection of the Cepheid classification, in particular for those located in the outer disc, will indicate whether this feature is real or simply an artefact of the inclusion of type~II Cepheids belonging to the thick disc in the current samples.
}


   \maketitle


\section{Introduction}

\par Type~II Cepheids are the older, fainter, low-mass counterpart to the classical Cepheids \citep[e.g.,][]{Wall2002}. As such, they fall into the instability strip between the RR~Lyrae and the RV~Tau stars, and their periods are bound to $\approx$~1 day on the lower end and to $\approx$~20 days on the upper end, following the classification of \citet{Sos2008b}. However, the limit between RR~Lyrae and Type~II Cepheids, on the one hand, and between Type~II Cepheids and RV~Tau stars, on the other, are not clearly defined. Type~II Cepheids are themselves divided in two sub-classes; the BL~Her stars have periods ranging from $\approx$~1 to $\approx$~4 days while the W~Vir stars have periods between $\approx$~4 and $\approx$~20 days, again according to \citet{Sos2008b}. In our current understanding, the different classes correspond to stars in different evolutionary stages: BL~Her stars are currently evolving from the horizontal branch (HB) to the asymptotic giant branch (AGB) and can be considered as post early-AGB stars \citep{Cast2007}. W~Vir stars cross the instability strip in their so-called ``blue-nose'' from the AGB while they are undergoing He-shell flashes. Finally, RV~Tau stars are about to leave the AGB, so they are crossing the instability strip towards the white dwarf domain \citep[][and references therein; see also \cite{Maas2007} for further considerations]{Gin1985,Bono1997}. \\ 

\par Various criteria have been tested to distinguish between classical and Type~II Cepheids, and among the Type~II Cepheids, to distinguish the BL~Her and the W~Vir stars. They are based on the shape of the light curve, on the stability of the period, or on the presence of distinctive features in the spectra. If these criteria have proved to be useful \citep[see for instance][]{Schmidt2004a}, they are not sufficient to secure a robust classification. Indeed for various criteria, the properties of different types of variables overlap over various period ranges \citep[e.g.,][]{Schmidt2005b,Sos2008b}. Moreover, the use of external parameters (metallicity, proper motion, distance to the Galactic plane) is hampered by the fact that Type~II Cepheids are very heterogeneous, because they span a wide metallicity range, and they can be found in the bulge, the thick disc, the halo, or in globular clusters.\\

\begin{table*}[ht!]
\centering
\caption{Spectroscopic observations of HQ~Car and DD~Vel.}
\label{obslog}
\centering
\begin{tabular}{c c c c c c c c c c}
\hline\hline
  Star &      RA     &     Dec     &   V   &   Epoch  &  Period  &       JD      & Phase & Spectrograph \\
       &      dms    &     dms     &  mag  &     d    &     d    &       d       &       &              \\ 
\hline  
HQ Car & 10 20 32.00 & -61 14 57.4 & 11.84 & 2452784.603\tablefootmark{a} & 14.06378\tablefootmark{a} & 2450834.86370 & 0.361 &  CTIO\tablefootmark{c}\\
       &             &             &       &          &                           & 2455284.65800 & 0.766 &  FEROS \\
       &             &             &       &          &                           & 2456411.53798 & 0.891 &  HARPS \\
       &             &             &       &          &                           & 2456412.51417 & 0.961 &  HARPS \\
\hline
DD Vel & 09 12 09.63 & -50 22 33.6 & 12.18 & 2434746.312\tablefootmark{b} & 13.1948\tablefootmark{b}  & 2454186.09017 & 0.291 &  FEROS\tablefootmark{d} \\
       &             &             &       &          &                           & 2454186.11160 & 0.292 &  FEROS\tablefootmark{d} \\
       &             &             &       &          &                           & 2454186.13302 & 0.294 &  FEROS\tablefootmark{d} \\
       &             &             &       &          &                           & 2454186.15443 & 0.296 &  FEROS\tablefootmark{d} \\
       &             &             &       &          &                           & 2454773.35198 & 0.798 &  UVES  \\
\hline
\end{tabular}
\tablefoot{
\tablefoottext{a}{Computed from ASAS photometry}
\tablefoottext{b}{GCVS values}
\tablefoottext{c}{Spectrum used to derive the chemical composition of HQ~Car}
\tablefoottext{d}{Spectra coadded and used to derive the chemical composition of DD~Vel}
}
\end{table*}  

\par The difficulty for properly classifying the Type~II Cepheids can be illustrated by the two stars in our sample: DD~Vel and HQ~Car. DD~Vel is classified as a classical Cepheid pulsating in the fundamental mode in the ASAS catalogue \citep{Pojmanski2002} and in the Machine-learned ASAS Classification Catalog \citep[MACC,][]{Richards2012} but as a Type~II Cepheid in both the General Catalog of Variable Stars \citep[GCVS,][]{Samus2007} and in the International Variable Star Index \citep[VSX,][]{Watson2006}. It is not listed as a Type~II Cepheid by \cite{Harris1985}. HQ~Car is considered as a classical Cepheid in the GCVS and in both the ASAS and MACC catalogues, while it is listed as a Type~II Cepheid by \cite{Harris1985} and in the VSX. It is worth mentioning that neither of the two stars is listed in the Fernie database of classical Cepheids \citep{Fernie1995}.\\

\par Surprisingly, Type~II Cepheids did not receive much attention \citep[see references in][]{Maas2007} after the early spectroscopic analyses of \cite{RodBell1963,RodBell1968}, \cite{Bar1971}, and \cite{AndKr1971}. Indeed, modern high resolution spectroscopic studies have been limited to only a few stars: TX Del \citep{And2002}, ST Pup \citep{Gon1996}, V553 Cen \citep{Wall1996}, and RT Tra \citep{Wall2000} before \cite{Maas2007} analysed 19 BL~Her and W~Vir stars. The study of \cite{Maas2007} is, to date, the unique extensive abundance analysis of Type~II Cepheids. Amongst other results, they made three significant points regarding the classification of Type~II Cepheids:
\begin{itemize}
 \item Amongst stars with short periods, there is a clear separation into two groups. The BL~Her stars are relatively metal-rich and show excesses of sodium, carbon, and nitrogen, along with thick disc kinematics. The UY~Eri stars are significantly more metal-poor and are similar to stars in the halo;
 \item Stars with periods between 10 and 20 days, such as W~Vir itself, show metallicities ranging approximately from --1.0 to --2.0 and are similar to variables in globular clusters;
 \item Stars with periods longer than 20 days show element separation as do RV~Tau stars. A few stars with periods in the 20--30 day range, such as TW~Cap, are as metal-poor as the 10--20-day stars.
\end{itemize}
\indent Since the \cite{Maas2007} paper, only two new Type~II Cepheids have been studied in detail: QQ~Per by \cite{Wall2008} and W~Vir by \cite{Kov2011}.

\par After a brief description of the data in Sect.~\ref{data}, we examine in Sect.~\ref{class} different classification criteria and discuss the chemical composition of HQ~Car and DD~Vel in Sect.~\ref{chem}.


\section{Data}
\label{data}

\par The HQ~Car spectra have been obtained with different instruments: one spectrum was taken with the echelle spectrograph on the 4m Blanco telescope at Cerro Tololo Inter-American Observatory (CTIO). It has a resolution of 28~000 and covers the 5500--8000 \AA{} wavelength range with a S/N (per pixel) of 57 in the order containing H$\alpha$. Another spectrum was obtained using the 2.2m ESO/MPG telescope and the FEROS echelle spectrograph at the ESO La Silla observatory \citep{Kau1999}. The spectrum covers the 3500--9200 \AA{} wavelength range with a resolution of 48~000 and a S/N (per pixel) in excess of 150 over the largest part of the spectrum. Finally, two spectra were obtained with the HARPS \citep{Mayor2003} echelle spectrograph mounted at the 3.6m telescope at the ESO La Silla observatory, which provides a resolution R=115 000 over a wide spectral range (3800--6900 \AA). They both reach a S/N of 50 at 650 nm.

\par We analysed two spectra for DD~Vel: the first one consists of four back-to-back FEROS spectra\footnote{prog. ID: 060.A-9120} coadded in order to increase the S/N. The second\footnote{prog. ID: 082.D-0901} was obtained with the UVES \citep{Dekker2000} echelle spectrograph (R=40~000) using the DIC2 (437+760) configuration. The blue and red arms cover the wavelength intervals [3750--5000] \AA{} and [5650--7600/7660–9460]~\AA. Relevant information concerning the observations and pulsation parameters of the Cepheids are listed in Table~\ref{obslog}. As shown in the next sections, strong emission features become prominent at some phases, and the spectra are therefore not suitable for an accurate abundance determination. We used the CTIO spectrum ($\phi$=0.361) for HQ~Car and the FEROS spectrum ($\phi$$\approx$0.292) in the case of DD~Vel. 
       
        
\section{Classification}
\label{class}

\subsection{Classification based on the location on a colour-magnitude diagram}

\begin{figure}[ht!]
        \includegraphics[angle=0,width=\columnwidth]{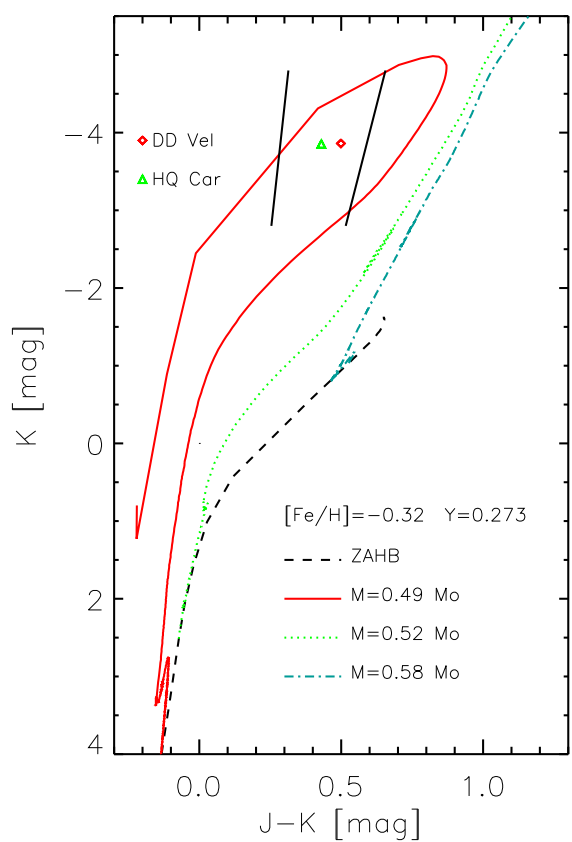}
        \caption{Location of DD~Vel and HQ~Car in a K,J-K colour-magnitude diagram. The dashed line shows the zero-age-horizontal-branch (ZAHB), while the coloured lines display HB evolutionary models for stellar masses ranging from 0.49 to 0.58 M$_{\odot}$. The black lines display the instability strip for RR~Lyrae and BL~Herculis stars.}
        \label{CMD}
\end{figure}

\par  Following the suggestion of an anonymous referee, we performed a detailed comparison in the K,J-K colour-magnitude diagram to constrain the nature of the candidate Type II Cepheids. Figure~\ref{CMD} shows evolutionary prescriptions for $\alpha$-enhanced horizontal branch (HB) evolutionary models \citep{Pietri2004,Pietri2006} at fixed chemical composition (see labelled values) and the two targets. It shows the Zero-Age-Horizontal-Branch (ZAHB )and HB evolutionary models for three different values of the stellar masses ranging from 0.49 to 0.58 M$_{\odot}$. The apparent NIR magnitudes of the targets are based on 2MASS photometry \citep{Skru2006}. They were unreddened using the empirical reddening law provided by \citet{Car1989}. The true distance modulus was estimated using the K-band period-luminosity relation for Type II Cepheids provided by \citet{Matsu2006}. We found M$_{K}$= --3.81 mag for DD~Vel and M$_{K}$ = --3.87 mag for HQ~Car. Data plotted in this figure show that the position of the targets agrees quite well, within the errors, with the current evolutionary prescriptions, thus further supporting the working hypothesis that they are Type~II Cepheids. It also displays the instability strip for RR~Lyrae and BL~Herculis stars. The hottest edge shows the first overtone blue edge, while the coolest shows the fundamental red edge. Current pulsation predictions (Marconi et al., 2015, ApJ, submitted) suggest that the edges of the instability strip are independent of the metal content in the NIR bands. The above edges should be cautiously treated, since they have been slightly extrapolated to higher luminosities to cover the magnitude range of the targets.

\subsection{Classification based on emission features in the spectrum}

\subsubsection{Emission in H$\alpha$}

The presence of emission in the H$\alpha$ lines of W~Vir was first reported by \cite{Joy1937} and \cite{Sanford1953}. \cite{Joy1949} noted that hydrogen line emission was a common feature of Type~II Cepheids in globular clusters; it has afterwards also been reported in field Type~II Cepheids \citep[e.g.,][]{Wall1958,Harris1984}. Following \cite{Schwarz1953}, the doubled absorption profiles, together with emission with an inverse P-Cygni profile in H$\alpha,$ were early modeled as a shock-wave passing through the atmosphere in the rising part of the lightcurve \citep{Whit1956a,Whit1956b,Wall1959,Whit1963}: emission originates in the de-excitation region behind the radiative shock wave. \cite{Lebre1992} used high resolution spectroscopy to follow the evolution of the emission profile of H$\alpha$ through an entire cycle of W~Vir and further improved the shock model. Finally, \cite{Kov2011} extended this study to many metallic lines including Fe~I, Fe~II, Na~I, and Ba~II and concluded that W~Vir consists in its inner part of a pulsating star with periodic shocks reaching the upper atmosphere and in its outer part of a circumstellar envelope.

\begin{figure}[htp]
        \begin{flushright}
        \includegraphics[angle=0,width=0.9\columnwidth]{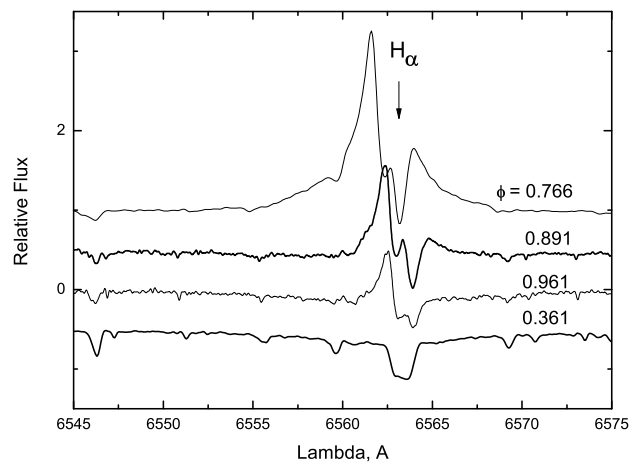} 
        \caption{Behaviour of the H$\alpha$ line in HQ~Car at different phases.}
        \label{Em_H_HQCar}
\end{flushright}
\end{figure}

\begin{figure}[h!]
        \begin{flushright}
        \includegraphics[angle=-90,width=0.88\columnwidth]{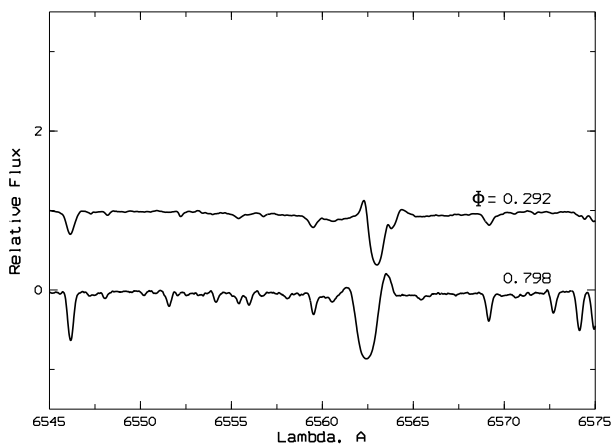}  
        \caption{Behaviour of the H$\alpha$ line in DD~Vel at different phases.}
        \label{Em_H_DDVel}
\end{flushright}
\end{figure}

In Fig.~\ref{Em_H_HQCar}, we present the variations in the H$\alpha$ profile for HQ~Car at four different phases. They are very similar to those presented for W~Vir by \citet[][their Fig. 3]{Lebre1992} and by \citet[][their Fig. 12]{Kov2011}. The shock wave rising in the atmosphere of the star causes a broad emission feature comprising five components (3 in emission, 2 in absorption). The absorption features are associated to the presence of a circumstellar envelope for one and to the fall back of the upper atmosphere located above the shock for the other. In particular, the H$\alpha$ emission totally disappears at {\it $\phi$}=0.361, in good agreement with \cite{Lebre1992}, who mention that the shock emission is present during the entire cycle except between phases {\it $\phi$}=0.38--0.44. In the case of DD~Vel (Fig~\ref{Em_H_DDVel}), the H$\alpha$ emission is weak as expected for {\it $\phi$}=0.292 but, more surprisingly, also weak at {\it $\phi$}=0.798. Given our limited phase coverage for this star, it could also very well be that we missed the phase of strong H$\alpha$ emission. 

\subsubsection{Emission in He~I at 5875.64 \AA}

\begin{figure}[htp]
        \includegraphics[angle=0,width=\columnwidth]{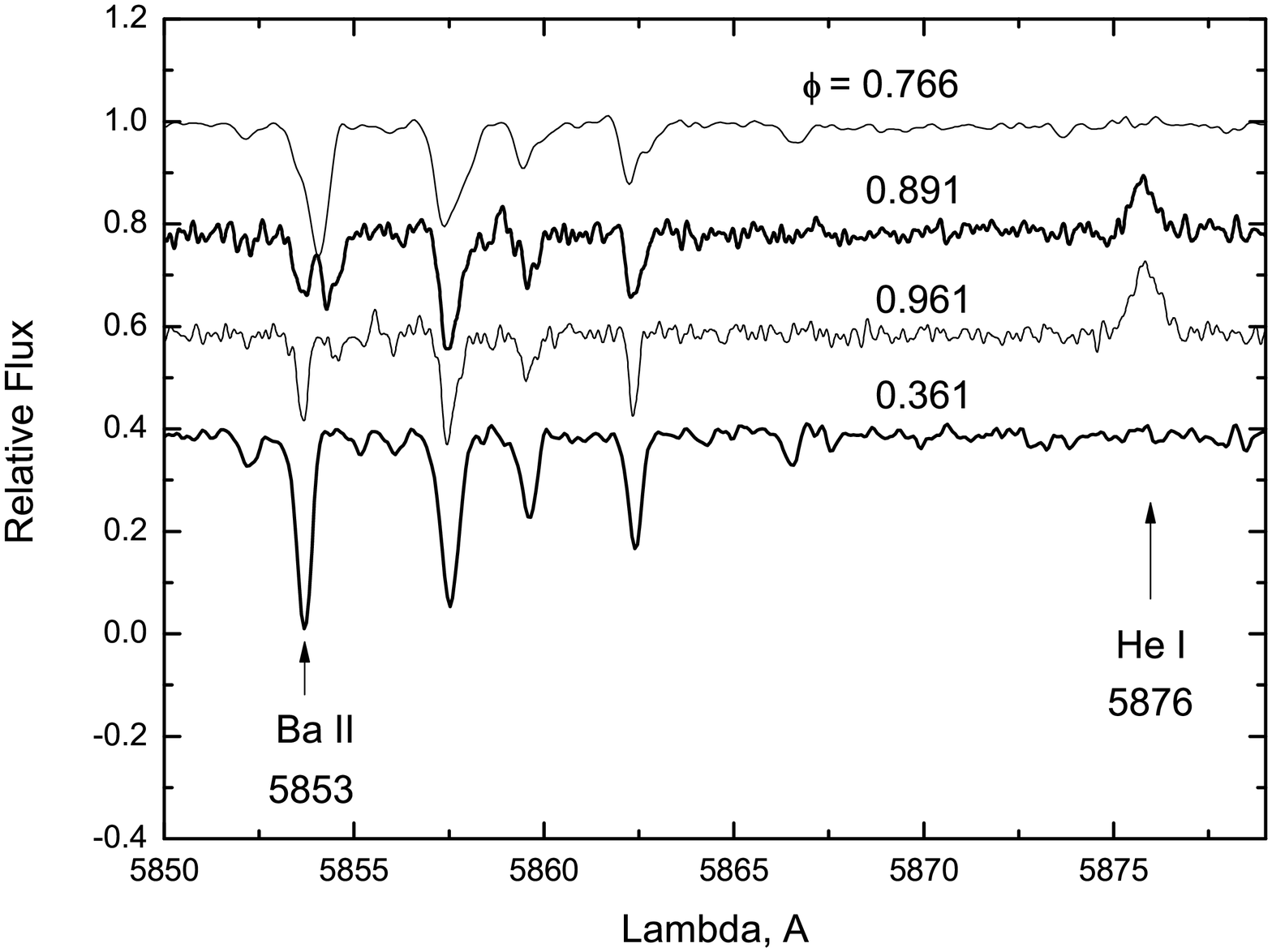}
        \caption{Behaviour of the 5853 Ba II line and the 5876 He I line in HQ~Car at different phases.}
        \label{Em_He_HQCar}
\end{figure}
\begin{figure}[hb!]
        \begin{flushright}
        \includegraphics[angle=-90,width=0.985\columnwidth]{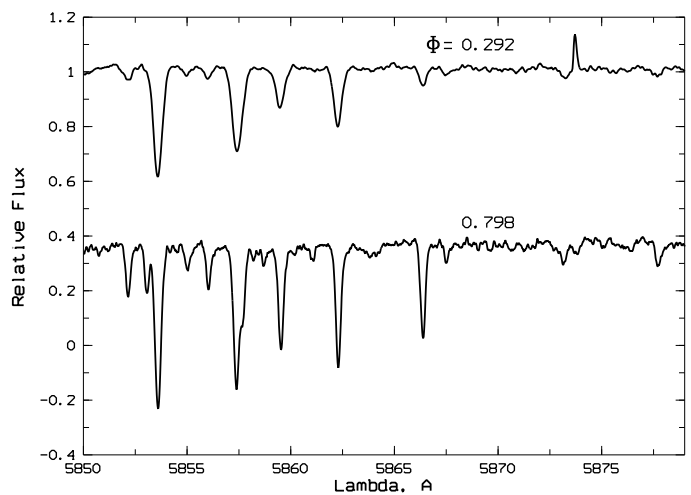}
        \caption{Same as Fig.~\ref{Em_He_HQCar} for DD~Vel.}
        \label{Em_He_DDVel}
\end{flushright}
\end{figure}

Emission lines of He~I in the spectra of type~II Cepheids were first mentioned by \cite{Wall1959}. They enabled him to confirm the shock model. (The emission is caused by helium ionized by the shock wave that captures electrons.) More recent observations by \cite{Raga1989} were used to determine the H/He ratio in the atmosphere of W~Vir. He~I emission lines were extensively studied by both \cite{Lebre1992} and \cite{Kov2011}. The former reported the presence of emission in the He 5875.64~\AA{} line between phases {\it $\phi$}=0.827 and {\it $\phi$}=0.009, while the latter detected emission between {\it $\phi$}=0.865 and {\it $\phi$}=0.201 for the same line. In both studies the emission peaks between {\it $\phi$}$\sim$0.8 and {\it $\phi$}$\sim$0.1, when the shock reaches its highest intensity.\\

In Fig.~\ref{Em_He_HQCar}, we present the variations in the He~I 5875.64\AA{} line profile for HQ~Car at four different phases. Once again, they are very similar to those presented for W~Vir by \citet[][their Fig. 6]{Lebre1992} and by \citet[][their Figs. 4,8]{Kov2011}. Emission is present only at the end of the cycle, at phases {\it $\phi$}=0.891 and {\it $\phi$}=0.961, in good agreement with previous studies. In the same figure, we note the doubling of the Ba~II line profile in the same phases. The mechanism responsible for the line doubling was first explained by \cite{Schwarz1953}: line doubling can be observed when the shock wave moves across the layer of formation of a given absorption line, provided that this layer is thick enough; the blueshifted line is produced by cooling gas moving upwards, while the redshifted line originates in gas already falling down. In the case of DD~Vel, Fig.~\ref{Em_He_DDVel} shows no emission for the He~I 5875.64\AA{} line, but the Ba~II line is split into two components at {\it $\phi$}=0.798, indicating that the region where the Ba~II line is formed is crossed by the shock wave.

\subsection{Kinematics consistent with a thick disc membership}

\par Kinematics alone is not sufficient to decide that a star belongs to the thin or the thick disc. It is, however, interesting to investigate the kinematic properties of DD~Vel and HQ~Car. Therefore we computed their space velocities U$_{LSR}$, V$_{LSR}$, W$_{LSR}$ in the local standard of rest\footnote{[U$_{\odot}$, V$_{\odot}$, W$_{\odot}$] = [11.10, 12.24, 7.25] km/s \citep{Schon2010}} using proper motions from the Naval Observatory Merged Astrometric Dataset \citep[NOMAD,][]{Zacharias2004} and the data shown in Table~\ref{kin}. Both stars have a total velocity 70~$\leq$~v$_{tot}$~$\leq$~180~km/s, making them likely thick disc members \citep[e.g.,][]{Nissen2004}. Their velocity along the direction of Galactic rotation V$_{LSR}$ almost falls within --100 km/s and --40 km/s, the range quoted by \cite{Reddy2003} for a probable thick-disc membership. Finally, comparing the stars in our sample to the velocity--metallicity plots of \citet[][see their Fig.~1]{Bensby2007}, we find that U$_{LSR}$ is not conclusive for HQ Car, while V$_{LSR}$ and W$_{LSR}$ both place this star in the thick disc. The situation is a bit less clear for DD~Vel because its U$_{LSR}$ is still at the upper limit for a thin disc star, and its V$_{LSR}$ at the lower limit, whereas its W$_{LSR}$ is typical of the thick disc. In conclusion, HQ~Car seems to be a very likely thick-disc member ,whereas it cannot be totally excluded that DD~Vel is a thin-disc member. The above kinematical evidence therefore supports the hypothesis that both stars are Type~II Cepheids in the thick disc rather than classical Cepheids located in the thin disc.

\begin{table*}[ht!]
\centering
\caption{Kinematics of HQ~Car and DD~Vel.}
\label{kin}
\centering
\begin{tabular}{c c c r r c c r r r r}
\hline\hline
  Star &      RA     &     Dec     &   PM (RA)   &   PM (Dec)  &  Vrad  & Distance &U$_{LSR}$&V$_{LSR}$&W$_{LSR}$&v$_{LSR}$\tablefootmark{e}\\
       &      deg    &     deg     &   mas/yr~~  &   mas/yr~~  &  km/s  &    pc    & km/s & km/s & km/s & km/s\\ 
\hline  
HQ Car & 155.133 & --61.249 & 1.1$\pm$2.6\tablefootmark{a} & 2.9$\pm$2.6\tablefootmark{a} & 62.05\tablefootmark{b} & 5725\tablefootmark{d} & 10.13 & --57.07 & 85.55 & 103.34\\
\hline
DD Vel & 138.040 & --50.376 & --5.7$\pm$4.7\tablefootmark{a} & --1.3$\pm$4.7\tablefootmark{a} & 26.02\tablefootmark{c} & 2444\tablefootmark{d} &  44.78  & --33.23 &  --51.75 & 76.08 \\
\hline
\end{tabular}
\tablefoot{
\tablefoottext{a}{NOMAD catalogue \citep{Zacharias2004}.}
\tablefoottext{b}{Radial velocity measured in our CTIO spectrum. The other spectra show line doubling with radial velocities of 85.26 \& 108.7 km/s (FEROS), 79.78 \& 113.8 km/s, 74.68 \& 120.0 km/s (HARPS), respectively.
}
\tablefoottext{c}{$\gamma$-velocity of DD~Vel from \cite{Metzger1992}.}
\tablefoottext{d}{Distance derived from the apparent magnitude in the J band (2MASS) and the period-luminosity relation of \cite{Matsu2006}.}
\tablefoottext{e}{ v$_{LSR}$ = (U$^{2}$$_{LSR}$+V$^{2}$$_{LSR}$+W$^{2}$$_{LSR}$)$^{1/2}$}
}
\end{table*}  

\subsection{Classification based on the chemical composition}

\par From their period (with respect to the classification of \citet{Sos2008b} for Type~II Cepheids) and the emission features in H$\alpha$ and He~I at 5876~\AA, it already appears to be clear that both HQ~Car and DD~Vel are W~Vir stars. This will be reinforced in Sect.~\ref{chem} where we examine their chemical composition.


\section{Chemical composition}
\label{chem}

\subsection{Method}

\par We used the DECH 30 software package\footnote{http://www.gazinur.com/DECH-software.html} to normalize the individual spectra to the local continuum, to identify the lines of different chemical elements, and to measure the equivalent widths (EW) of the absorption lines. The oscillator strengths have been taken from the Vienna Atomic Lines Database \citep[VALD][]{Kup1999}.

\par To determine the effective temperature T$_{\rm eff}$, we employed the line depth ratios method of \cite{Kov2007}, which comes from the work of \cite{KovGor2000}. The ratios of the central depths of carefully chosen pairs of lines that have a very different dependence on T$_{\rm eff}$ are entered in previously calibrated relations. This technique allows determining T$_{\rm eff}$ with great precision: the use of several tens ($\geq$50) of ratios per spectrum leads to uncertainties of $\approx$10-20~K when S/N$>$100 and of $\approx$30-50~K when S/N$<$100. The method is independent of the interstellar reddening and only marginally dependent on the individual characteristics of stars, such as rotation, microturbulence, and metallicity.\\ 

\par To determine the surface gravity ($log~g$) and the microturbulent velocity V$_{t}$, we used a canonical analysis. We sought the surface gravity from the excitation equilibrium of Fe~I and Fe~II lines, and the microturbulent velocity is determined from the Fe~I lines. We note that the excitation equilibrium is also satisfied by V~I and V~II and, to a slightly lesser extent, by Ti~I and Ti~II in HQ~Car, while it is satisfied for the couples Si~I~/~Si~II, Ti~I~/~Ti~II (but not Cr~I~/~Cr~II) in the case of DD~Vel. As far as the microturbulent velocity is concerned, an innovative approach using lines of several elements has been developed by \cite{Sahin2011} and is illustrated in \cite{Reddy2012}. In this method, the standard errors are plotted as a function of the microturbulent velocity. We applied it to the stars in our sample, and the results are in good agreement with our values for V$_{t}$. They are described in Appendix \ref{App_vt}. The atmospheric parameters for DD~Vel and HQ~Car are listed in Table~\ref{atmparam}.

\begin{table}[ht]
\centering\small
\caption{Atmospheric parameters derived for HQ Car and DD Vel.}
\begin{tabular}{rcccc}
\hline
\hline
Star & $T_{\rm eff}$ & $\log~g$ & $V_{\rm t}$ & [Fe/H] \\
     &     K         &    dex   &     km/s    & dex \\  
\hline
HQ~Car & 5580 & 1.6 & 3.1 & --0.32 \\  
DD~Vel & 5572 & 1.4 & 3.8 & --0.48 \\    
\hline
\end{tabular}
\label{atmparam}
\end{table}

\par The lines of odd-Z elements can be broadened due to their hyperfine structure (hfs). However, the hfs corrections are negligible in the case of V or Co for the considered EW. This is not true in the case of Sc, Mn, or Cu \citep[e.g.,][]{North2012,Reddy2012}. We therefore computed the abundances of these elements via spectral synthesis using the 5526.79, 5657.90, 5667.15, 6245.62, 6604.60 lines for Sc~II, 5420.35, 5432.56, 6013.48, 6021.79 for Mn~I and
5105.55, 5218.21, 5782.14 for Cu~I, and the STARSP code developed by \cite{Tsym1996}. We took the hyperfine structure of Sc~II \citep{Pro2000}, Mn~I, and Cu~I \citep{Allen2011} into account for the line profile calculations.

\par Atmospheric models are interpolated for each Type~II Cepheid using the grid of 1D, LTE atmosphere models of \cite{Cast2004}. Individual abundances are listed in Table~\ref{AbH} and abundance ratios (with respect to iron) in Table~\ref{AbFe}. We computed the solar reference abundances using lines in the Sun with EWs $<$ 120mA and the same atmosphere models \citep{Cast2004}. They are listed in the Appendix~\ref{App_solar}, together with the prescriptions of \cite{Asp2009} and the solar abundances of \cite{Reddy2003} that are used by \cite{Reddy2006} in their study of the thick disc.\\

\begin{table}
\centering\small
\caption[]{Individual abundances [X/H] in HQ Car and DD Vel.}
\begin{tabular}{rr|rcc|rcc}
\hline\hline
\multicolumn{2}{c}{} & \multicolumn{3}{c}{HQ Car}&\multicolumn{3}{c}{DD Vel}\\
\hline
\multicolumn{2}{c|}{Ion}&[X/H]&$\sigma$& N &[X/H]&$\sigma$& N \\
\hline
   C {\sc i}  &  6.00 & --0.14 & 0.19 &   7  & --0.48 & 0.15 &   2 \\
   N {\sc i}  &  7.00 &   0.20 & 0.21 &   2  &   0.23 & 0.11 &   2 \\
   O {\sc i}  &  8.00 &   0.23 & 0.01 &   2  & --0.22 &      &   1 \\
  Na {\sc i}  & 11.00 & --0.25 & 0.09 &   3  & --0.33 & 0.12 &   4 \\
  Mg {\sc i}  & 12.00 & --0.17 &      &   1  & --0.41 &      &   1 \\
  Al {\sc i}  & 13.00 & --0.43 & 0.18 &   4  &        &      &     \\
  Si {\sc i}  & 14.00 & --0.11 & 0.12 &  22  & --0.24 & 0.14 &  20 \\
  Si {\sc ii} & 14.01 &        &      &      & --0.28 & 0.15 &   2 \\
   S {\sc i}  & 16.00 & --0.05 & 0.12 &   5  & --0.39 & 0.00 &   2 \\
  Ca {\sc i}  & 20.00 & --0.47 & 0.18 &   9  & --0.73 & 0.06 &  14 \\
  Sc {\sc ii} & 21.01 & --0.55\tablefootmark{1} &      &      & --1.12\tablefootmark{1} &      &     \\
  Ti {\sc i}  & 22.00 & --0.35 & 0.19 &   6  & --0.61 & 0.11 &  13 \\
  Ti {\sc ii} & 22.01 & --0.28 &      &   1  & --0.61 & 0.11 &   4 \\
   V {\sc i}  & 23.00 & --0.39 & 0.13 &   9  & --0.48 & 0.10 &   4 \\
   V {\sc ii} & 23.01 & --0.43 & 0.13 &   2  &        &      &     \\
  Cr {\sc i}  & 24.00 & --0.59 & 0.23 &   3  & --0.85 & 0.06 &   7 \\
  Cr {\sc ii} & 24.01 &        &      &      & --0.51 & 0.10 &   7 \\
  Mn {\sc i}  & 25.00 & --0.63\tablefootmark{1} &      &      & --0.71\tablefootmark{1} &      &     \\
  Fe {\sc i}  & 26.00 & --0.32 & 0.10 & 133  & --0.48 & 0.11 & 162 \\
  Fe {\sc ii} & 26.01 & --0.32 & 0.07 &   8  & --0.51 & 0.09 &  19 \\
  Co {\sc i}  & 27.00 & --0.18 & 0.19 &   5  & --0.38 & 0.17 &   3 \\
  Ni {\sc i}  & 28.00 & --0.34 & 0.13 &  38  & --0.52 & 0.06 &  44 \\
  Cu {\sc i}  & 29.00 & --0.48\tablefootmark{1} &      &      & --0.43\tablefootmark{1} &      &     \\
  Zn {\sc i}  & 30.00 &        &      &      & --0.20 & 0.05 &   2 \\
   Y {\sc ii} & 39.01 & --0.82 & 0.10 &   2  & --1.41 & 0.13 &   3 \\
  Zr {\sc ii} & 40.01 & --1.16 &      &   1  &        &      &     \\
  La {\sc ii} & 57.01 & --0.72 & 0.08 &   3  &        &      &     \\
  Nd {\sc ii} & 60.01 & --0.58 &      &   1  & --0.89 & 0.07 &   2 \\
  Eu {\sc ii} & 63.01 & --0.04 &      &   1  &        &      &     \\
\hline
\end{tabular}
\tablefoot{
\tablefoottext{1}{Abundance determined by spectral synthesis}
}
\label{AbH}
\end{table}

\begin{table}
\centering\small
\caption[]{Abundance ratios [X/Fe] in HQ~Car and DD~Vel.}
\begin{tabular}{rr|rc|rc|rc}
\hline\hline
\multicolumn{2}{c}{} & \multicolumn{4}{c}{HQ~Car}&\multicolumn{2}{c}{DD~Vel}\\
\hline
\multicolumn{2}{c|}{Ion}&[X/Fe]&$\sigma$&[X/Fe]\tablefootmark{1}&$\sigma$\tablefootmark{1}\\
\hline
    C {\sc i}  &  6.00 &  +0.18 & 0.21 &        &      &  +0.00 & 0.19 \\
    N {\sc i}  &  7.00 &  +0.52 & 0.23 &        &      &  +0.71 & 0.16 \\
    O {\sc i}  &  8.00 &  +0.55 & 0.10 &        &      &  +0.26 & 0.11 \\
   Na {\sc i}  & 11.00 &  +0.07 & 0.13 &        &      &  +0.15 & 0.16 \\
   Mg {\sc i}  & 12.00 &  +0.15 & 0.10 &        &      &  +0.07 & 0.11 \\
   Al {\sc i}  & 13.00 & --0.11 & 0.21 &        &      &        &      \\
   Si {\sc i}  & 14.00 &  +0.21 & 0.16 &  +0.44 & 0.18 &  +0.24 & 0.18 \\
   Si {\sc ii} & 14.01 &        &      &        &      &  +0.20 & 0.19 \\
    S {\sc i}  & 16.00 &  +0.27 & 0.16 &        &      &  +0.09 & 0.11 \\
   Ca {\sc i}  & 20.00 & --0.15 & 0.21 & --0.04 & 0.22 & --0.25 & 0.13 \\
   Sc {\sc ii} & 21.01 & --0.23 & 0.10 &        &      & --0.64 & 0.11 \\
   Ti {\sc i}  & 22.00 & --0.03 & 0.21 & --0.01 &      & --0.13 & 0.16 \\
   Ti {\sc ii} & 22.01 &  +0.04 & 0.10 & --0.07 &      & --0.13 & 0.16 \\
    V {\sc i}  & 23.00 & --0.07 & 0.16 &        &      &  +0.00 & 0.15 \\
    V {\sc ii} & 23.01 & --0.11 & 0.16 &        &      &        &      \\
   Cr {\sc i}  & 24.00 & --0.27 & 0.25 &        &      & --0.37 & 0.13 \\
   Cr {\sc ii} & 24.01 &        &      &        &      & --0.03 & 0.15 \\
   Mn {\sc i}  & 25.00 & --0.31 & 0.10 &        &      & --0.23 & 0.11 \\
   Co {\sc i}  & 27.00 &  +0.14 & 0.21 &        &      &  +0.10 & 0.20 \\
   Ni {\sc i}  & 28.00 & --0.02 & 0.16 &        &      & --0.04 & 0.13 \\
   Cu {\sc i}  & 29.00 & --0.16 & 0.10 &        &      &  +0.05 & 0.11 \\
   Zn {\sc i}  & 30.00 &        &      &        &      &  +0.28 & 0.12 \\
    Y {\sc ii} & 39.01 & --0.50 & 0.14 &        &      & --0.93 & 0.17 \\
   Zr {\sc ii} & 40.01 & --0.84 & 0.10 &        &      &        &      \\
   La {\sc ii} & 57.01 & --0.40 & 0.13 &        &      &        &      \\
   Nd {\sc ii} & 60.01 & --0.26 & 0.10 &        &      & --0.41 & 0.13 \\
   Eu {\sc ii} & 63.01 &  +0.28 & 0.10 &  +0.06 &      &        &      \\
\hline
\end{tabular}
\tablefoot{
\tablefoottext{1}{\cite{Yong2006}}
}
\label{AbFe}
\end{table}

\par We used 25 calibrations to determine the effective temperature of HQ~Car and 26 calibrations for DD~Vel, leading to standard deviations of 95~K and 109~K, respectively, and standard errors of 19~K and 22~K. We adopted 100~K as the uncertainty on $T_{\rm eff}$. We estimated the uncertainty on $log~g$ as $\pm$0.2~dex and the uncertainty on $V_{\rm t}$ as $\pm$0.5~km/s. Table \ref{unc} lists the variations in the individual abundances [X/H] when changing the atmospheric parameters by $\Delta$$T_{\rm eff}$=+100~K, $\Delta$$\log~g$=+0.2~dex, and $\Delta$$V_{\rm t}$=+0.2~km/s and their sum in quadrature, which we adopt as the uncertainty on the abundances due to the uncertainties on the atmospheric parameters. It is well documented \citep[e.g.,][]{John2002} that such a method leads to overestimated values for the total error, because by construction it ignores covariances between the different atmospheric parameters. They nevertheless remain lower than 0.10 dex in most cases. The sum in quadrature of the errors associated with the uncertainties on the atmospheric parameters and of the standard deviation associated with the determination of the abundance of a given element gives the total error on the abundance for this element.

\begin{table}
\centering\small
\caption{Abundance uncertainties due to uncertainties on the atmospheric parameters, computed for HQ~Car ($T_{\rm eff}$=5580~K, $\log~g$=1.6 dex, $V_{\rm t}$=3.1 km/s, [Fe/H]=--0.327 dex). Cols. 3--5 are the errors associated with the uncertainties on one of the individual atmospheric parameters. Col. 6 is the total error associated with the uncertainties on the atmospheric parameters. Col. 7 is the total error, adding in quadrature col. 6 and the standard deviations listed in Table \ref{AbH} for individual species.}
\label{unc}
\begin{tabular}{rrccccc}
\hline\hline
\multicolumn{2}{c|}{Ion}& $\Delta$ $T_{\rm eff}$ & $\Delta$ $\log~g$ & $\Delta$ $V_{\rm t}$ & Total & Total \\
\multicolumn{2}{c|}{}  &          +100~K        &    +0.2~dex       &      +0.5~km/s        & (atm)  &        \\ 
\hline
C  {\sc  i}  &   6.00 & --0.06&    0.08&  --0.02&  0.10 &  0.21 \\
N  {\sc  i}  &   7.00 & --0.09&    0.08&  --0.02&  0.12 &  0.24 \\
O  {\sc  i}  &   8.00 &   0.05&    0.08&  --0.03&  0.10 &  0.10 \\
Na {\sc  i}  &  11.00 &   0.05&    0.00&  --0.04&  0.06 &  0.11 \\
Mg {\sc  i}  &  12.00 &   0.05&    0.00&  --0.07&  0.09 &  0.09 \\
Al {\sc  i}  &  13.00 &   0.03&  --0.01&  --0.01&  0.03 &  0.18 \\
Si {\sc  i}  &  14.00 &   0.04&    0.00&  --0.03&  0.05 &  0.13 \\
Si {\sc ii}  &  14.01 & --0.07&    0.09&  --0.12&  0.17 &  0.17 \\
S  {\sc  i}  &  16.00 & --0.03&    0.07&  --0.03&  0.08 &  0.14 \\
Ca {\sc  i}  &  20.00 &   0.07&    0.00&  --0.09&  0.11 &  0.21 \\
Sc {\sc ii}  &  21.01 &   0.03&    0.08&  --0.08&  0.12 &  0.12 \\
Ti {\sc  i}  &  22.00 &   0.09&    0.00&  --0.02&  0.09 &  0.21 \\
Ti {\sc ii}  &  22.01 &   0.02&    0.07&  --0.04&  0.08 &  0.08 \\
V  {\sc  i}  &  23.00 &   0.10&  --0.01&  --0.01&  0.10 &  0.16 \\
V  {\sc ii}  &  23.01 &   0.01&    0.07&  --0.02&  0.07 &  0.15 \\
Cr {\sc  i}  &  24.00 &   0.05&  --0.01&  --0.01&  0.05 &  0.24 \\
Mn {\sc  i}  &  25.00 &   0.07&  --0.01&  --0.05&  0.09 &  0.09 \\
Fe {\sc  i}  &  26.00 &   0.07&  --0.01&  --0.05&  0.09 &  0.13 \\
Fe {\sc ii}  &  26.01 & --0.01&    0.08&  --0.04&  0.09 &  0.11 \\
Co {\sc  i}  &  27.00 &   0.11&  --0.01&  --0.01&  0.11 &  0.22 \\
Ni {\sc  i}  &  28.00 &   0.07&  --0.01&  --0.04&  0.08 &  0.15 \\
Cu {\sc  i}  &  29.00 &   0.10&  --0.01&  --0.03&  0.10 &  0.10 \\
Y  {\sc ii}  &  39.01 &   0.02&    0.07&  --0.01&  0.07 &  0.12 \\
Zr {\sc ii}  &  40.01 &   0.02&    0.07&  --0.01&  0.07 &  0.07 \\
La {\sc ii}  &  57.01 &   0.05&    0.07&  --0.01&  0.09 &  0.12 \\
Nd {\sc ii}  &  60.01 &   0.05&    0.07&    0.00&  0.09 &  0.09 \\
Eu {\sc ii}  &  63.01 &   0.04&    0.08&  --0.04&  0.10 &  0.10 \\
  \hline
\end{tabular}
\end{table}
          
\subsection{Chemical composition}
             
\par The two stars in our sample have [Fe/H] in the --0.3 to --0.5 dex range, towards the lower end of the metallicity distribution for BL~Her stars, but still in a domain where the metallicities of BL~Her and W~Vir stars overlap \citep[see][]{Maas2007}. As shown just after this, they are probably affected by dust-gas separation. However, the [S/Fe] we measured for DD~Vel (+0.09 dex) and HQ~Car (+0.27~dex) are very similar to those already reported for [S/Fe] in different Galactic structures. Below [Fe/H]=--1.0 dex, [S/Fe], values are scattered around a plateau at $\approx$+0.25 dex and decrease at higher metallicities until reaching [S/Fe]=0.0 dex at [Fe/H]=--0.3 dex \citep[e.g.,][and references therein]{Fran1987,Fran1988,Chen2002,Nissen2007,Mat2013,Caffau2014}. Also our [Zn/Fe] measurement of +0.28$\pm$0.12 dex in DD~Vel is very consistent with previous values ([Zn/Fe]$\approx$+0.1--+0.2~dex) reported for the thick disc \citep[e.g.,][]{Mish2002,Bens2005,Brew2006,Reddy2006}. Since sulphur and zinc are only slightly depleted into dust \citep{Savage1996}, the typical thick-disc values for [S/Fe] and [Zn/Fe] in HQ~Car and DD~Vel indicate that their iron abundances are probably not very modified by the dust-gas separation. In particular, this allows us to use an average thick disc star for the --0.45 to --0.55 dex [Fe/H] bin for comparison purposes \citep[][col. 4 in their Table 7]{Reddy2006}.\\
             
\par
\begin{figure}[h!]
        \includegraphics[angle=-90,width=\columnwidth]{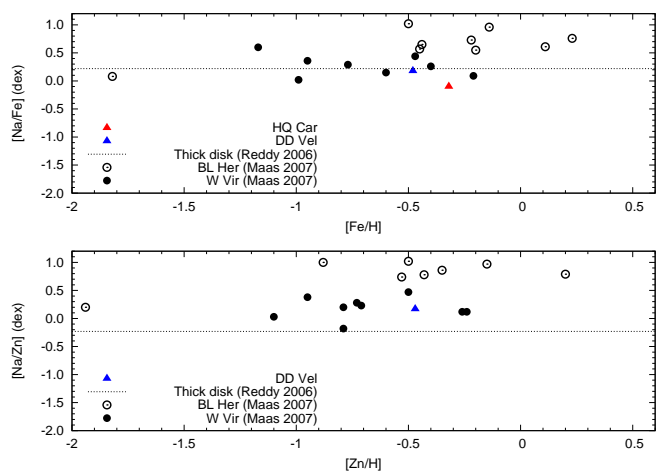}
        \caption{{\it Top:} [Na/Fe] vs. [Fe/H] for BL~Her stars (open circles) and W~Vir stars (filled circles) from \cite{Maas2007}. HQ~Car and DD~Vel are overplotted in red and blue,  respectively. The ratio for a representative thick-disc star at [Fe/H]=--0.5 dex is shown as a dashed line. {\it Bottom:}  same for [Na/Zn] vs. [Zn/H]. Abundances have been rescaled to the solar abundances of \cite{Reddy2003}, used as a reference by \cite{Maas2007}.}
        \label{Na}
\end{figure}

 \cite{Maas2007} have shown that W~Vir stars have [Na/Fe] that is independent of [Fe/H] and consistent with thick disc stars where <[Na/Fe]>=+0.12 dex \citep{Reddy2006}. This argument does not apply to the stars with a severe dust-gas separation. They find, in contrast, that BL~Her stars are strongly overabundant in sodium with a mean [Na/Fe]=+0.73 dex. For the two Type~II Cepheids in our sample, [Na/Fe] varies between +0.07 and +0.15 dex, similar to the representative thick disc value (see Fig.~\ref{Na}). Since iron could be affected by dust-gas separation, we also compare [Na/Zn] for the BL~Her and W~Vir stars, and again DD~Vel has low [Na/Zn] similar to the other W~Vir stars, while the BL~Her stars show very high ($>$+0.5 dex) values of [Na/Zn]. The absence of Na overabundance in our sample confirms that they are W~Vir stars and not BL~Her stars, as could already be inferred from their period (P$>$4d).\\    
             
\par
\begin{figure*}[ht!]
\centering   
\begin{minipage}[b]{.5\textwidth}
  \centering 
  \includegraphics[width=.99\textwidth]{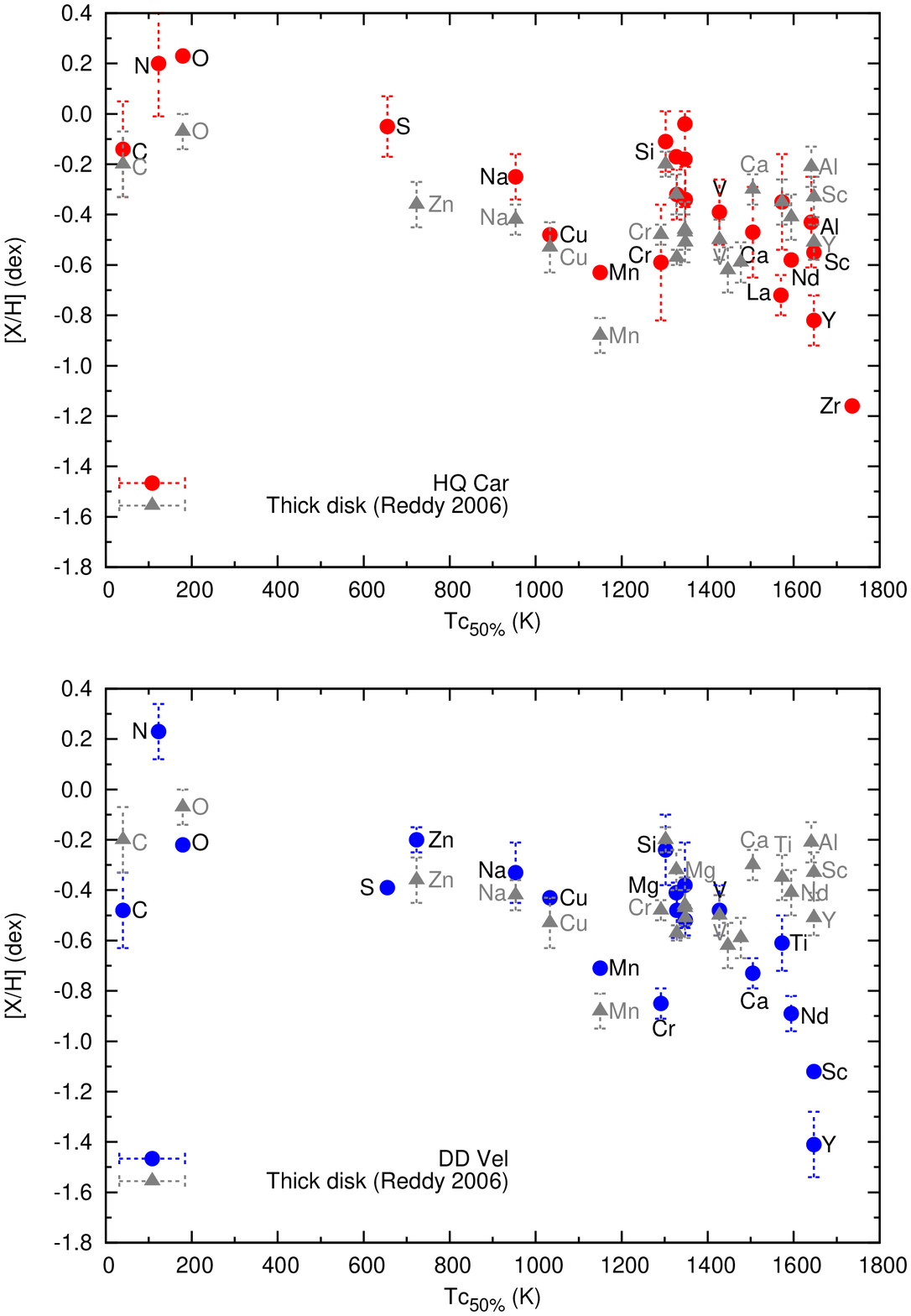}
  \captionof{figure}{[X/H] vs. 50\%~T$_{c}$ for the two Type~II Cepheids in our sample. An average thick-disc star (grey dots) is plotted for comparison. Elements are identified by their chemical symbol. Abundances have been rescaled to the solar abundances that we recomputed with the \cite{Cast2004} models (See Appendix~\ref{App_solar})}
  \label{abTc}
\end{minipage}%
\begin{minipage}[b]{.5\textwidth}
  \centering
  \includegraphics[width=.99\textwidth]{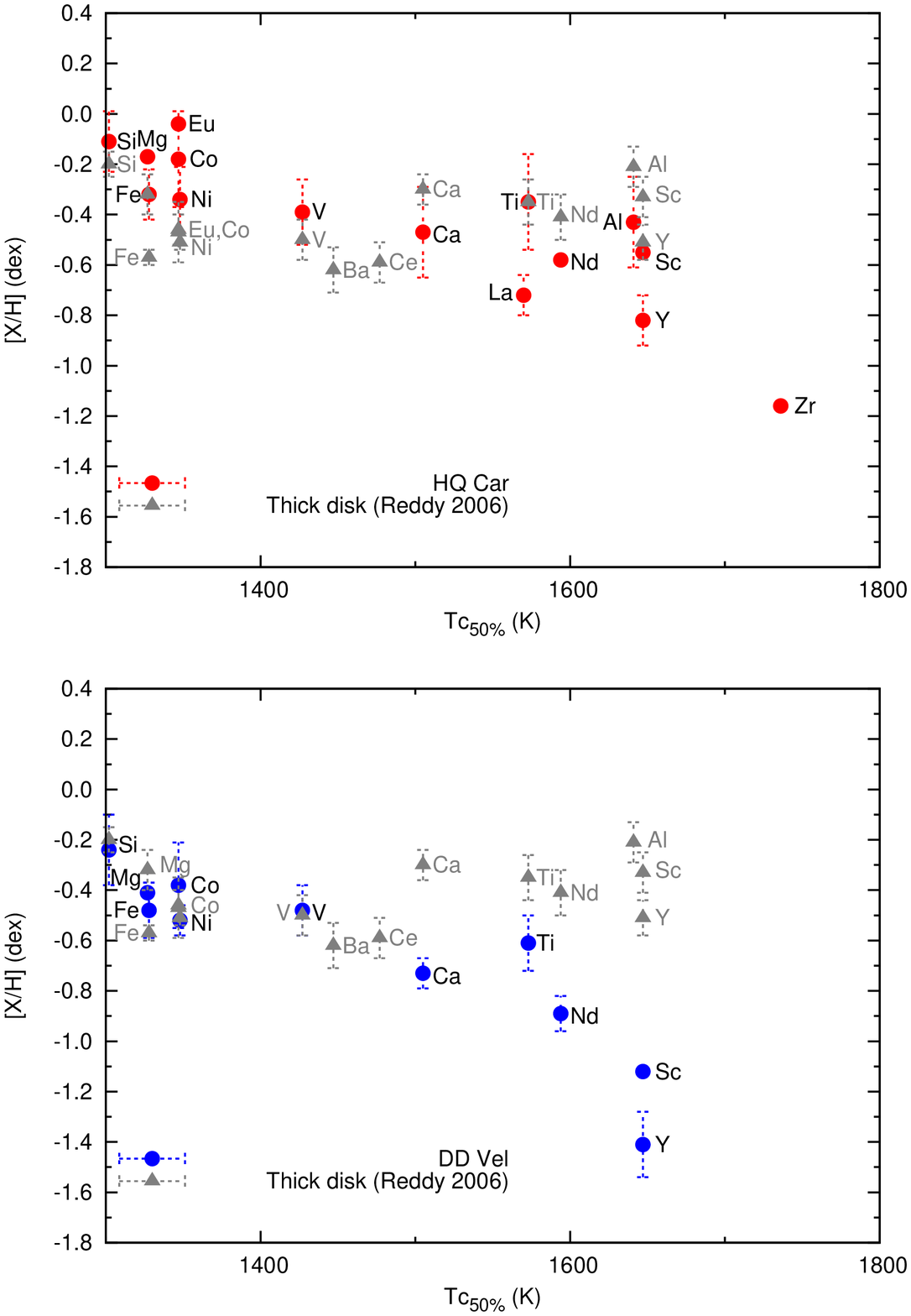}
  \captionof{figure}{Same as Fig.~\ref{abTc}, but focusing on the elements with the highest condensation temperature.\newline \newline \newline \newline}
  \label{abTczoom}
\end{minipage}
\end{figure*}

 When gas cools sufficiently, dust grains can form and the abundances of the elements in the gas phase decrease. Because this happens at different temperatures for different trace elements, the quantity "50\% condensation temperature" (50\%~T$_{c}$) has been defined, at which 50\% of the element is found in the gas phase and the other 50\% is locked in dust grains. We adopted the 50\%~T$_{c}$ determined by \cite{Lodd2003}. A correlation between the underabundance of a given element and its condensation temperature is then interpreted as a dust-gas separation\footnote{It should be noted that this assumption involves implicit approximations (such as the formation of dust grains under conditions of thermodynamic equilibrium) that may not be met in the surroundings of W~Vir stars due to the presence of shocks in the atmosphere \citep{Kov2011}.}. As can be seen in Fig.~\ref{abTc}, the Type~II Cepheids in our sample show hints of (mild) dust-gas separation: the more volatile elements have abundances similar to those of an average thick-disc star \citep{Reddy2006}, while the refractory elements are underabundant, because they are depleted into dust. For a better visibility, we focus in Fig.~\ref{abTczoom} on the elements with 50\%~T$_{c}$~$>$~1300~K.

\par As expected, the signature of dust-gas separation is especially marked for the elements with the highest 50\%~T$_{c}$. In HQ~Car, Ca, Nd, Al, and Sc are mildly depleted by $\approx$--0.2 dex and Y by $\approx$--0.3 dex with respect to an average thick-disc star. Zr seems to be very depleted because [Zr/H] clearly falls below the abundances of the other neutron-capture elements, but we have no comparison with an average thick-disc star to draw a firm conclusion. The depletion is more severe for DD~Vel as the underabundances with respect to an average thick disc star reach $\approx$--0.30 dex for Ti,  $\approx$--0.45 dex for Nd and Ca, $\approx$--0.80 dex for Sc and $\approx$--0.90 dex for Y. To further support the dust-gas separation, we note that most of the individual abundances fall below those of the thick-disc reference star for the elements with 50\%~T$_{c}$~$>$~1400~K. 

\par Hints or even clear evidence of dust-gas separation in Type~II Cepheids have already been reported for ST~Pup by \cite{Gon1996} and for CO~Pup, V1711~Sgr, MZ~Cyg, and SZ~Mon by \cite{Maas2007}. On the other hand, these authors find the signature of dust-gas separation less convincing in RX~Lib and W~Vir because it relies mostly on the depletion in Sc. As we discuss in more detail in the next paragraph, severe dust-gas separation has also been reported in most of the RV~Tau stars \citep[see][and references therein]{Giri2005}. \cite{Maas2007} found a Type~II Cepheid (CC~Lyr) with an extreme dust-gas separation, which is larger than in any RV~Tau star. It is important to note that all the Type~II Cepheids with a signature of dust-gas separation are W~Vir stars and not BL~Her stars. That the stars in our sample also show possible (HQ~Car) or probable (DD~Vel) signs of this phenomenon reinforces their classification as W~Vir stars.\\ 

\par Dust-gas separation is a common feature in RV~Tau stars \citep[see][and references therein]{Giri2005}. In a qualitative scenario \citep{Waters1992}, binary RV~Tau stars are surrounded by a dusty disc. The dust-gas separation occurs when radiation pressure traps the dust grains in the disc while some of the gas (deprived from dust) is re-accreted on the star via the viscous disc that allows for transfer of angular momentum. The origin of the circumbinary disc in RV~Tau stars is not clear, but it is generally believed to be created during binary interaction when the primary was a giant.

\par This scenario excludes metal-poor systems ([Fe/H]$\leq$--1.0 dex) where dust cannot form in sufficient quantities, and indeed no sign of dust-gas separation has been found for metal-poor RV~Tau variables in globular clusters \citep{Gon1997}. Similarly, \cite{Maas2007} find no evidence of dust-gas separation for TW~Cap, a W~Vir star with [Fe/H]=--1.8 dex possibly associated to the halo of the Milky Way.\\

\par It is not clear that the dust-gas separation has the same origin in W~Vir stars as in RV~Tau stars. In particular, the observed depletion is generally much shallower in the W~Vir stars. Only in the case of CC~Lyr does it reach the extreme values more commonly seen in RV~Tau stars \citep[e.g., $\approx$--3.0 dex in HP~Lyr and DY~Ori, see][]{Giri2005}. The RV~Tau stars depleted in their refractory elements are known binaries for a large number of them, supporting the hypothesis of a circumbinary dusty disc \citep{Rao2012}. Disentangling orbital velocities from pulsational velocities is very demanding in terms of observing time both in the case of RV~Tau and W~Vir stars, and indeed only four type~II Cepheids are currently known as binaries: AU~Peg, IX~Cas, TX~Del, and ST~Pup. It is interesting to note that the only W~Vir star in this group (ST~Pup) shows obvious signs of dust-gas depletion, while the other shorter-period stars do not.\\

\par Recently reported observational \citep[e.g.,][]{Mar2010} and theoretical \citep[e.g.,][]{Neil2012} lines of evidence support the existence of mass loss in classical Cepheids; however, these outflows seem to have a very low dust content \citep{Mar2013}, possibly indicating that the wind is driven by pulsation and is not dust-driven as generally observed in evolved stars. On the other hand, extended dusty environments have been detected with high angular resolution techniques \citep[e.g.,][and references therein]{Ker2006,Gal2013} and from extended emission in the mid- and far-infrared \citep{Bar2011}. They have been attributed to the presence of a circumstellar envelope around the Cepheids.
\par As far as Type~II Cepheids are concerned, \cite{Kov2011} analysed hydrogen, helium, and metallic lines in W~Vir itself. They were able to reproduce the specifics of spectral line variability in W~Vir with the help of a non-linear pulsation model. Results suggest that W~Vir consists of two different layers, the inner part being the pulsating star itself and the outer part a very extended and dense atmosphere, and it might even include a circumstellar envelope with a very low expansion rate. 
\par If the most desirable experiment in the near future were to systematically examine the binarity properties of W~Vir stars, it would nevertheless be interesting to investigate whether dust-gas separation could also somehow take place in their circumstellar envelopes.

               
\section{Summary and conclusion.}

\par The status of the HQ~Car and DD~Vel Type~II Cepheids has remained unclear. Depending on the catalogue (i.e., on the method and criteria used to perform the classification), they are sometimes listed as classical Cepheids and sometimes as Type~II. Because we observed emission features in the H$\alpha$ and in the 5875.64 \AA{} He~I lines that are characteristic features of W~Vir stars, we conclude that HQ~Car and DD~Vel are Type~II Cepheids from this sub-class. Their periods of 14.06 and 13.19 days, respectively, and the absence of Na overabundance further indicates that they are not BL~Her stars. Moreover, they show a possible (HQ~Car) or probable (DD~Vel) signature of mild dust-gas separation. Such abundance patterns have currently been observed only in long-period Type~II Cepheids and RV~Tau stars, thus reinforcing our classification.

\par Several studies of the Galactic abundance gradients in the thin disc using classical Cepheids have reported increased dispersion in the outer disc \citep{Yong2006,Lem2008,Luck2011a,Luck2011b,Gen2014}; however, these findings are hampered by the possible contamination of the current samples by misclassified Type~II Cepheids that are thick-disc members. Including unrecognized Type~II Cepheids modifies the abundance patterns not only because they belong to another stellar population but also because their current distances are computed with period-luminosity relations that are only valid for classical Cepheids. Thick-disc members can be identified by their specific location in the [$\alpha$/Fe] vs [Fe/H] plane \citep[e.g.,][]{Recio2014}, and recent studies indicate that the thin disc contamination by thick-disc stars is not negligible \citep[see, for instance,][their Fig.~7]{Miko2014}.


\begin{acknowledgements}
The authors thank the anonymous referees for the very valuable comments that helped to improve the quality of this paper. 
The authors thank Dr. D. Yong and Prof. B.W. Carney for providing the HQ Car spectrum. 
The authors thank Profs. C. Dominik and L.B.F.M Waters for useful discussions on the dust-gas separation phenomenon.
V.K.  acknowledges the support from the Swiss National Science Foundation,
project SCOPES No. IZ73Z0$_{}$152485. GG acknowledges the support of the Chilean fund FONDECYT-regular (project 1120190). 
\end{acknowledgements}
\vspace{0.5cm}


\onecolumn
\newpage
\begin{appendix} 

\section{An alternative approach to determining V$_{t}$}
\label{App_vt}

\par In this Appendix we show the standard deviation around the mean abundance plotted as a function of the microturbulent velocity, following the approach of \cite{Sahin2011}. The dispersion of the abundances is computed for the Fe I, Fe II, Si I, and Ni I lines, while the microturbulent velocity V$_{t}$ is varied from 1 to 6 km/s. The minimum value of the dispersion is in good agreement within the different elements, confirming the values of the microturbulence (derived solely from Fe I) adopted in this study, namely 3.1 km/s for HQ~Car and 3.8 km/s for DD~Vel.. 

\begin{figure*}[ht!]
\centering   
\begin{minipage}[b]{.5\textwidth}
  \centering 
  \includegraphics[width=.9\textwidth]{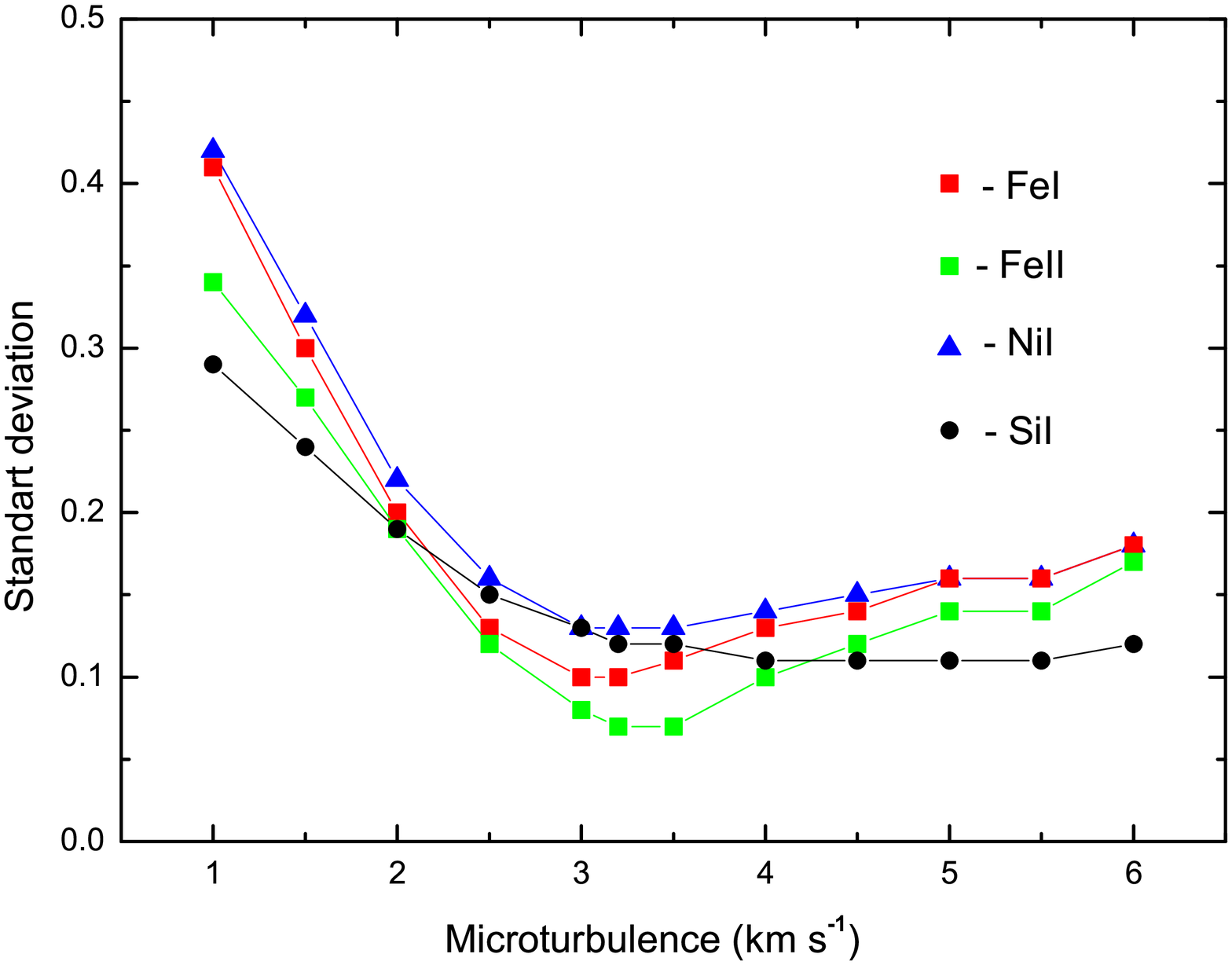}
  \captionof{figure}{Standard deviation around the mean abundances as a function of microturbulence V$_{t}$ for HQ Car, shown for several elements.}
  \label{vt_HQCar}
\end{minipage}%
\begin{minipage}[b]{.5\textwidth}
  \centering
  \includegraphics[width=.9\textwidth]{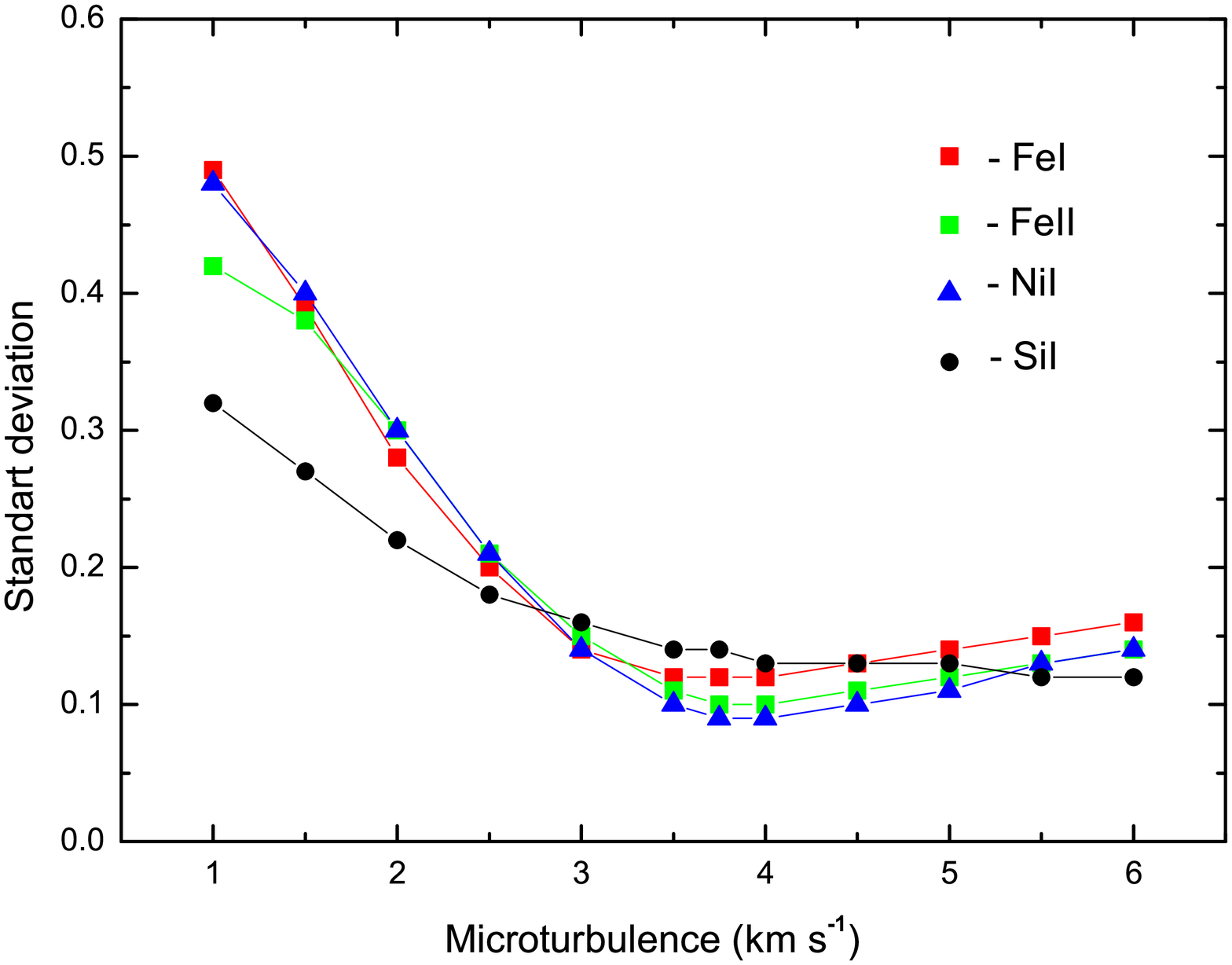}
  \captionof{figure}{Same as Fig.~\ref{vt_HQCar}, but for DD Vel.\newline \newline}
  \label{vt_DDVel}
\end{minipage}
\end{figure*}

\vspace{-0.5 cm}
\section{Solar references}
\label{App_solar}
\vspace{-0.5 cm}
 
\begin{table}[ht!]
\centering\small
\caption{Solar abundance derived from the solar spectrum using the grid of models by \cite{Cast2004}, compared to the solar photospheric abundance by \cite{Asp2009} and by \cite{Reddy2003}.}
\label{solar}
\begin{tabular}{llrll}
\hline\hline
 Element       & Log A             & Lines &        Log A          &          Log A          \\
               &   (this work)     & used  &    \citep{Asp2009}    &   \citep{Reddy2003}     \\
               &        dex        &       &         dex           &        dex              \\ 
\hline
    C {\sc i}  &    8.53$\pm$0.14  &   4   &      8.43$\pm$0.05    &    8.51$\pm$0.06        \\
    N {\sc i}  &    8.15           &   2   &      7.83$\pm$0.05    &    8.06                 \\
    O {\sc i}  &    8.98           &   1   &      8.69$\pm$0.05    &    8.86$\pm$0.05        \\
   Na {\sc i}  &    6.32$\pm$0.04  &  10   &      6.24$\pm$0.04    &    6.27                 \\
   Mg {\sc i}  &    7.68$\pm$0.02  &   9   &      7.60$\pm$0.04    &    7.54$\pm$0.06        \\
   Al {\sc i}  &    6.30$\pm$0.01  &   2   &      6.45$\pm$0.03    &    6.28$\pm$0.05        \\
   Si {\sc i}  &    7.55$\pm$0.08  &  23   &      7.51$\pm$0.03    &    7.62$\pm$0.05        \\
    S {\sc i}  &    7.20$\pm$0.10  &   6   &      7.12$\pm$0.03    &    7.34$\pm$0.09        \\    
   Ca {\sc i}  &    6.32$\pm$0.07  &  16   &      6.34$\pm$0.04    &    6.33$\pm$0.07        \\
   Sc {\sc i}  &                   &       &      3.15$\pm$0.04    &                         \\
   Sc {\sc ii} &    3.22$\pm$0.11  &  14   &                       &    3.24$\pm$0.14        \\
   Ti {\sc i}  &    4.96$\pm$0.08  &  41   &      4.95$\pm$0.05    &    4.90$\pm$0.06        \\
   Ti {\sc ii} &    5.01$\pm$0.03  &   5   &                       &                         \\
    V {\sc i}  &    4.04$\pm$0.12  &  36   &      3.93$\pm$0.08    &    3.93$\pm$0.03        \\
   Cr {\sc i}  &    5.67$\pm$0.09  &  23   &      5.64$\pm$0.04    &    5.68$\pm$0.07        \\
   Mn {\sc i}  &    5.54$\pm$0.07  &  11   &      5.43$\pm$0.05    &    5.37$\pm$0.05        \\
   Fe {\sc i}  &    7.57$\pm$0.08  & 164   &      7.50$\pm$0.04    &    7.45$\pm$0.04        \\
   Fe {\sc ii} &    7.47$\pm$0.04  &  11   &                       &    7.45$\pm$0.07        \\
   Co {\sc i}  &    5.00$\pm$0.10  &  28   &      4.99$\pm$0.07    &    4.93$\pm$0.04        \\
   Ni {\sc i}  &    6.29$\pm$0.06  &  56   &      6.22$\pm$0.04    &    6.23$\pm$0.04        \\
   Cu {\sc i}  &    4.29$\pm$0.20  &   3   &      4.19$\pm$0.04    &    4.19$\pm$0.05        \\
   Zn {\sc i}  &    4.45           &   1   &      4.56$\pm$0.05    &    4.47                 \\
    Y {\sc ii} &    2.15$\pm$0.17  &   7   &      2.21$\pm$0.05    &    2.12$\pm$0.04        \\
   Zr {\sc i}  &                   &       &      2.58$\pm$0.04    &                         \\
   Zr {\sc ii} &    2.79$\pm$0.19  &   2   &                       &    2.45                 \\
   La {\sc ii} &    1.24$\pm$0.02  &   2   &      1.10$\pm$0.04    &                         \\
   Ce {\sc ii} &    1.70$\pm$0.11  &   6   &      1.58$\pm$0.04    &    1.58                 \\
   Nd {\sc ii} &    1.54$\pm$0.08  &  11   &      1.42$\pm$0.04    &    1.50                 \\
   Eu {\sc ii} &    0.96           &   1   &      0.52$\pm$0.04    &    0.61                 \\
\hline                                                        
\end{tabular}                                                 
\end{table}                                                   
                                                         
\end{appendix}                                                
 
\Online
\onecolumn

\begin{appendix}
\section{List of lines used}
\label{App_linelist}

\begin{longtable}{rrrrrr}
\caption{\label{EW} Atomic parameters and EWs of HQ~Car and DD~Vel}\\
\hline\hline
Wavelength &   Z   & Log gf & $\chi_{ex}$ &  HQ Car &  DD Vel  \\
 $\AA$     &       &        &     eV      & EW(m\AA)& EW(m\AA) \\ 
\hline
\endfirsthead
\caption{continued.}\\
\hline\hline
Wavelength &   Z   & Log gf & $\chi_{ex}$ &  HQ Car &  DD Vel  \\
 $\AA$     &       &        &     eV      & EW(m\AA)& EW(m\AA) \\ 
\hline
\endhead
\hline
\endfoot
   6587.6100 &   6.00&  -1.002 &  8.537&   58.2 &  33.6  \\
   7087.8300 &   6.00&  -1.441 &  8.647&   27.8 &   ...  \\
   7111.4700 &   6.00&  -1.084 &  8.640&   26.6 &  34.6  \\
   7113.1800 &   6.00&  -0.772 &  8.647&   51.6 &   ...  \\
   7115.1700 &   6.00&  -0.933 &  8.643&   51.6 &   ...  \\
   7116.9900 &   6.00&  -0.906 &  8.647&   71.4 &   ...  \\
   7119.6600 &   6.00&  -1.147 &  8.643&   49.8 &   ...  \\
   7442.2980 &   7.00&  -0.400 & 10.330&   13.8 &  24.8  \\
   7468.3120 &   7.00&  -0.182 & 10.336&   31.2 &  26.6  \\
   6300.3040 &   8.00&  -9.818 &  0.000&  107.9 &   ...  \\
   6363.7760 &   8.00&  -10.302&  0.020&   52.2 &  27.1  \\
   5682.6330 &  11.00&  -0.705 &  2.102&  105.8 &  97.6  \\
   5688.2050 &  11.00&  -0.451 &  2.104&    ... & 141.9  \\
   6154.2260 &  11.00&  -1.546 &  2.102&   32.6 &  29.0  \\
   6160.7470 &  11.00&  -1.245 &  2.104&   45.6 &  38.1  \\
   5711.0880 &  12.00&  -1.723 &  4.346&  132.2 & 123.4  \\
   6696.0230 &  13.00&  -1.346 &  3.143&   27.4 &   ...  \\
   6698.6730 &  13.00&  -1.646 &  3.143&    8.4 &   ...  \\
   7835.3090 &  13.00&  -0.648 &  4.022&   26.0 &   ...  \\
   7836.1340 &  13.00&  -0.493 &  4.022&   18.5 &   ...  \\
   5645.6130 &  14.00&  -2.139 &  4.930&   52.7 &  38.9  \\
   5665.5540 &  14.00&  -2.039 &  4.920&   53.4 &  37.7  \\
   5684.4840 &  14.00&  -1.649 &  4.954&    ... & 102.8  \\
   5690.4250 &  14.00&  -1.869 &  4.930&   81.8 &  51.5  \\
   5708.4000 &  14.00&  -1.469 &  4.954&    ... & 110.1  \\
   5772.1460 &  14.00&  -1.749 &  5.082&    ... &  60.8  \\
   5793.0730 &  14.00&  -2.059 &  4.930&   63.4 &  64.5  \\
   5948.5410 &  14.00&  -1.229 &  5.082&    ... & 114.2  \\
   6091.9190 &  14.00&  -1.462 &  5.871&   25.4 &  30.6  \\
   6106.6080 &  14.00&  -1.896 &  5.614&   21.2 &   ...  \\
   6125.0210 &  14.00&  -1.464 &  5.614&   44.3 &  31.2  \\
   6131.8520 &  14.00&  -1.616 &  5.616&   41.6 &   ...  \\
   6145.0160 &  14.00&  -1.310 &  5.616&   46.6 &  37.6  \\
   6155.1340 &  14.00&  -0.754 &  5.619&  117.4 &   ...  \\
   6237.3190 &  14.00&  -0.974 &  5.614&   87.2 &  62.4  \\
   6243.8150 &  14.00&  -1.243 &  5.616&   71.9 &  44.6  \\
   6244.4650 &  14.00&  -1.090 &  5.616&   56.8 &   ...  \\
   6414.9800 &  14.00&  -1.035 &  5.871&   66.5 &  45.9  \\
   6526.6300 &  14.00&  -1.606 &  5.871&    ... &  22.3  \\
   6721.8480 &  14.00&  -1.526 &  5.863&    ... &  37.4  \\
   6800.5960 &  14.00&  -1.944 &  5.964&   14.3 &   ...  \\
   6848.5800 &  14.00&  -1.527 &  5.863&   21.5 &   ...  \\
   7034.9010 &  14.00&  -0.879 &  5.871&    ... &  68.5  \\
   7226.2080 &  14.00&  -1.509 &  5.614&    ... &  31.3  \\
   7282.8160 &  14.00&  -0.625 &  6.206&   69.9 &  78.6  \\
   7373.0040 &  14.00&  -1.179 &  5.984&   34.0 &   ...  \\
   7424.6100 &  14.00&  -1.609 &  5.619&   24.0 &   ...  \\
   7680.2660 &  14.00&  -0.689 &  5.863&    ... &  95.5  \\
   7918.3830 &  14.00&  -0.609 &  5.954&  100.9 &   ...  \\
   7932.3480 &  14.00&  -0.469 &  5.964&  107.0 &   ...  \\
   7944.0010 &  14.00&  -0.309 &  5.984&  128.8 &   ...  \\
   6347.1090 &  14.01&   0.170 &  8.121&  166.4 & 132.4  \\
   6371.3710 &  14.01&  -0.039 &  8.121&    ... &  93.5  \\
   6045.9720 &  16.00&  -1.317 &  7.868&   44.8 &  29.0  \\
   6045.9920 &  16.00&  -1.907 &  7.868&   44.8 &  29.0  \\
   6046.0380 &  16.00&  -1.113 &  7.868&   44.8 &  29.0  \\
   6743.5800 &  16.00&  -0.849 &  7.866&   52.5 &   ...  \\
   6748.7900 &  16.00&  -0.529 &  7.868&   68.8 &  33.9  \\
   6757.1500 &  16.00&  -0.239 &  7.870&   83.6 &  41.4  \\
   7686.1010 &  16.00&  -0.987 &  7.868&   26.8 &   ...  \\
   5349.4650 &  20.00&  -0.309 &  2.709&    ... & 106.3  \\
   5512.9800 &  20.00&  -0.463 &  2.933&    ... &  60.7  \\
   5581.9650 &  20.00&  -0.554 &  2.523&  104.4 &  91.8  \\
   5590.1140 &  20.00&  -0.570 &  2.521&   97.3 &  88.4  \\
   5601.2770 &  20.00&  -0.522 &  2.526&  133.2 & 104.2  \\
   5867.5620 &  20.00&  -1.569 &  2.933&   22.4 &   ...  \\
   6161.2970 &  20.00&  -1.265 &  2.523&   42.9 &  37.7  \\
   6166.4390 &  20.00&  -1.141 &  2.521&   52.8 &  43.5  \\
   6169.0420 &  20.00&  -0.796 &  2.523&    ... &  74.2  \\
   6169.5630 &  20.00&  -0.477 &  2.526&  106.2 & 103.6  \\
   6449.8080 &  20.00&  -0.501 &  2.521&    ... & 111.2  \\
   6471.6620 &  20.00&  -0.685 &  2.526&  122.6 &  90.0  \\
   6493.7810 &  20.00&  -0.108 &  2.521&        & 162.1  \\
   6499.6500 &  20.00&  -0.817 &  2.523&   86.3 &  77.9  \\
   6717.6810 &  20.00&  -0.523 &  2.709&    ... & 105.9  \\
   7326.1450 &  20.00&  -0.207 &  2.933&    ... &  98.5  \\
   5020.0260 &  22.00&  -0.413 &  0.836&    ... &  94.1  \\
   5022.8680 &  22.00&  -0.433 &  0.826&    ... &  80.9  \\
   5024.8440 &  22.00&  -0.601 &  0.818&    ... &  68.5  \\
   5036.4640 &  22.00&   0.130 &  1.443&    ... &  90.0  \\
   5039.9570 &  22.00&  -1.089 &  0.021&    ... &  83.0  \\
   5210.3850 &  22.00&  -0.849 &  0.048&    ... & 110.8  \\
   5224.3000 &  22.00&   0.130 &  2.134&    ... &  37.2  \\
   5514.5330 &  22.00&  -0.359 &  1.443&    ... &  38.0  \\
   5644.1330 &  22.00&   0.231 &  2.267&   50.2 &  31.8  \\
   5866.4510 &  22.00&  -0.839 &  1.067&   35.7 &  33.1  \\
   6098.6580 &  22.00&  -0.009 &  3.062&    9.7 &   ...  \\
   6126.2160 &  22.00&  -1.424 &  1.067&   16.0 &   ...  \\
   6258.1020 &  22.00&  -0.354 &  1.443&    ... &  43.2  \\
   6258.7060 &  22.00&  -0.239 &  1.460&   59.7 &  51.3  \\
   6261.0970 &  22.00&  -0.478 &  1.430&   44.7 &  36.8  \\
   5005.1570 &  22.01&  -2.729 &  1.566&    ... &  98.3  \\
   5268.6150 &  22.01&  -1.669 &  2.598&    ... & 109.8  \\
   5381.0150 &  22.01&  -1.969 &  1.566&    ... &        \\
   5396.2260 &  22.01&  -3.019 &  1.584&    ... &  52.3  \\
   5418.7510 &  22.01&  -2.109 &  1.582&    ... &        \\
   6606.9500 &  22.01&  -2.789 &  2.061&   63.8 &  39.9  \\
   5670.8530 &  23.00&  -0.419 &  1.081&   19.7 &   ...  \\
   5698.5190 &  23.00&  -0.110 &  1.064&   35.1 &  30.4  \\
   5703.5750 &  23.00&  -0.210 &  1.051&   23.2 &   ...  \\
   5727.0480 &  23.00&  -0.011 &  1.081&   25.8 &  29.2  \\
   5731.2410 &  23.00&  -0.729 &  1.064&   14.0 &   ...  \\
   6090.2140 &  23.00&  -0.061 &  1.081&    ... &  33.4  \\
   6111.6450 &  23.00&  -0.714 &  1.043&   10.7 &   ...  \\
   6199.1970 &  23.00&  -1.299 &  0.287&   19.9 &  19.8  \\
   6216.3540 &  23.00&  -1.289 &  0.275&   23.8 &   ...  \\
   6242.8290 &  23.00&  -1.549 &  0.262&   13.1 &   ...  \\
   5819.9250 &  23.01&  -1.692 &  2.522&   29.8 &   ...  \\
   5928.8520 &  23.01&  -1.682 &  2.522&   42.8 &   ...  \\
   5247.5650 &  24.00&  -1.589 &  0.961&    ... &  91.8  \\
   5296.6910 &  24.00&  -1.359 &  0.983&    ... & 116.8  \\
   5297.3770 &  24.00&   0.209 &  2.900&    ... &  67.0  \\
   5329.1380 &  24.00&  -0.007 &  2.914&    ... &  50.3  \\
   5348.3150 &  24.00&  -1.209 &  1.004&    ... & 125.1  \\
   5783.8500 &  24.00&  -0.360 &  3.322&   18.6 &   ...  \\
   5787.9180 &  24.00&  -0.049 &  3.322&   28.6 &  29.1  \\
   6661.0750 &  24.00&  -0.112 &  4.193&   11.7 &   ...  \\
   7400.2490 &  24.00&  -0.049 &  2.900&        &  64.2  \\
   5237.3280 &  24.01&  -1.349 &  4.073&    ... & 160.9  \\
   5310.6860 &  24.01&  -2.407 &  4.072&    ... &  46.2  \\
   5313.5630 &  24.01&  -1.778 &  4.074&    ... & 115.3  \\
   5334.8690 &  24.01&  -1.825 &  4.072&    ... & 107.4  \\
   5407.6040 &  24.01&  -2.150 &  3.827&    ... &  82.5  \\
   5502.0670 &  24.01&  -2.116 &  4.168&    ... &  53.3  \\
   5508.6060 &  24.01&  -2.251 &  4.156&    ... &  56.1  \\
   5002.7930 &  26.00&  -1.579 &  3.397&    ... & 107.3  \\
   5004.0440 &  26.00&  -1.399 &  4.209&    ... &  50.3  \\
   5022.2360 &  26.00&  -0.529 &  3.984&    ... & 142.9  \\
   5029.6180 &  26.00&  -2.049 &  3.415&    ... &  53.9  \\
   5044.2110 &  26.00&  -2.037 &  2.851&    ... &  91.5  \\
   5048.4360 &  26.00&  -1.029 &  3.960&    ... &  83.1  \\
   5054.6430 &  26.00&  -1.920 &  3.640&    ... &  48.8  \\
   5060.0360 &  26.00&  -1.147 &  4.301&    ... &  74.6  \\
   5060.0790 &  26.00&  -5.459 &  0.000&    ... &  74.6  \\
   5067.1500 &  26.00&  -0.969 &  4.220&    ... &  84.8  \\
   5090.7740 &  26.00&  -0.399 &  4.256&    ... & 128.9  \\
   5109.6520 &  26.00&  -0.979 &  4.301&    ... &  94.6  \\
   5131.4690 &  26.00&  -2.514 &  2.223&    ... & 141.0  \\
   5159.0580 &  26.00&  -0.819 &  4.283&    ... &  89.6  \\
   5180.0700 &  26.00&  -1.259 &  4.473&    ... &  40.1  \\
   5187.9140 &  26.00&  -1.370 &  4.143&    ... &  49.3  \\
   5225.5260 &  26.00&  -4.788 &  0.110&    ... & 132.8  \\
   5228.3770 &  26.00&  -1.289 &  4.220&    ... &  58.7  \\
   5242.4910 &  26.00&  -0.966 &  3.634&    ... & 120.9  \\
   5243.7770 &  26.00&  -1.149 &  4.256&    ... &  65.9  \\
   5247.0500 &  26.00&  -4.945 &  0.087&    ... & 130.4  \\
   5253.4620 &  26.00&  -1.572 &  3.283&    ... &  99.4  \\
   5288.5250 &  26.00&  -1.507 &  3.695&    ... &  56.8  \\
   5307.3610 &  26.00&  -2.986 &  1.608&    ... & 151.2  \\
   5322.0410 &  26.00&  -2.802 &  2.279&    ... &  78.3  \\
   5329.9890 &  26.00&  -1.188 &  4.076&    ... &  58.6  \\
   5353.3740 &  26.00&  -0.839 &  4.103&    ... & 121.8  \\
   5373.7090 &  26.00&  -0.859 &  4.473&    ... &  66.3  \\
   5379.5740 &  26.00&  -1.513 &  3.695&    ... &  59.2  \\
   5389.4790 &  26.00&  -0.409 &  4.415&    ... & 110.1  \\
   5398.2790 &  26.00&  -0.669 &  4.446&    ... &  82.3  \\
   5409.1340 &  26.00&  -1.299 &  4.371&    ... &  52.4  \\
   5441.3390 &  26.00&  -1.729 &  4.313&    ... &  20.7  \\
   5464.2800 &  26.00&  -1.401 &  4.143&    ... &  34.2  \\
   5466.3960 &  26.00&  -0.629 &  4.371&    ... &  95.5  \\
   5473.9010 &  26.00&  -0.759 &  4.154&    ... & 100.1  \\
   5481.2430 &  26.00&  -1.242 &  4.103&    ... &  77.5  \\
   5487.7450 &  26.00&  -0.316 &  4.320&    ... & 129.9  \\
   5522.4470 &  26.00&  -1.549 &  4.209&    ... &  35.4  \\
   5543.1500 &  26.00&  -1.569 &  3.695&        &  66.6  \\
   5543.9360 &  26.00&  -1.139 &  4.218&   67.6 &  69.9  \\
   5546.5060 &  26.00&  -1.309 &  4.371&   56.7 &  51.1  \\
   5554.8950 &  26.00&  -0.439 &  4.549&  104.4 &   ...  \\
   5560.2120 &  26.00&  -1.189 &  4.435&   54.4 &  50.7  \\
   5563.6000 &  26.00&  -0.989 &  4.191&    ... &  99.9  \\
   5565.7040 &  26.00&  -0.212 &  4.608&  118.1 & 105.0  \\
   5567.3910 &  26.00&  -2.563 &  2.609&   83.1 &  67.9  \\
   5576.0890 &  26.00&  -0.999 &  3.430&  147.5 &   ...  \\
   5584.7650 &  26.00&  -2.319 &  3.573&   26.1 &  29.3  \\
   5618.6330 &  26.00&  -1.275 &  4.209&   51.9 &  42.7  \\
   5619.5950 &  26.00&  -1.699 &  4.387&   32.7 &   ...  \\
   5633.9470 &  26.00&  -0.269 &  4.991&   86.7 &  72.6  \\
   5638.2620 &  26.00&  -0.869 &  4.220&  101.3 &   ...  \\
   5650.7060 &  26.00&  -0.959 &  5.086&    ... &  30.5  \\
   5651.4690 &  26.00&  -1.999 &  4.473&   15.7 &   ...  \\
   5652.3180 &  26.00&  -1.949 &  4.260&   24.5 &   ...  \\
   5653.8670 &  26.00&  -1.639 &  4.387&   33.3 &  30.6  \\
   5661.3460 &  26.00&  -1.735 &  4.284&   18.7 &   ...  \\
   5662.5160 &  26.00&  -0.572 &  4.178&    ... & 126.5  \\
   5679.0230 &  26.00&  -0.919 &  4.652&   54.5 &  54.7  \\
   5686.5300 &  26.00&  -0.445 &  4.549&   96.4 &  80.6  \\
   5691.4970 &  26.00&  -1.519 &  4.301&   44.1 &  29.0  \\
   5701.5450 &  26.00&  -2.215 &  2.559&  131.4 & 126.4  \\
   5705.9920 &  26.00&  -0.529 &  4.608&    ... &  93.5  \\
   5717.8330 &  26.00&  -1.129 &  4.284&   76.0 &  64.0  \\
   5731.7620 &  26.00&  -1.299 &  4.256&   62.2 &  55.2  \\
   5741.8480 &  26.00&  -1.853 &  4.256&   19.4 &  23.1  \\
   5752.0320 &  26.00&  -1.176 &  4.549&   54.8 &  51.6  \\
   5753.1230 &  26.00&  -0.687 &  4.260&  108.3 & 100.0  \\
   5762.9920 &  26.00&  -0.449 &  4.209&    ... & 142.9  \\
   5775.0810 &  26.00&  -1.297 &  4.220&   61.7 &  56.6  \\
   5784.6580 &  26.00&  -2.531 &  3.397&    ... &  22.0  \\
   5793.9150 &  26.00&  -1.699 &  4.220&   24.6 &  34.5  \\
   5806.7250 &  26.00&  -1.049 &  4.608&   51.2 &  45.6  \\
   5809.2180 &  26.00&  -1.839 &  3.884&   43.5 &  44.2  \\
   5814.8080 &  26.00&  -1.969 &  4.283&   15.7 &   ...  \\
   5816.3740 &  26.00&  -0.600 &  4.549&   80.5 &   ...  \\
   5852.2190 &  26.00&  -1.329 &  4.549&   36.1 &  24.5  \\
   5859.5860 &  26.00&  -0.418 &  4.549&   89.2 &   ...  \\
   5862.3560 &  26.00&  -0.126 &  4.549&  122.6 & 109.6  \\
   5883.8170 &  26.00&  -1.359 &  3.960&   82.6 &  69.0  \\
   5905.6720 &  26.00&  -0.729 &  4.652&   58.4 &  53.6  \\
   5909.9740 &  26.00&  -2.586 &  3.211&    ... &  45.0  \\
   5916.2470 &  26.00&  -2.993 &  2.453&    ... &  67.9  \\
   5927.7890 &  26.00&  -1.089 &  4.652&   36.0 &   ...  \\
   5930.1800 &  26.00&  -0.229 &  4.652&  117.9 & 116.0  \\
   5934.6550 &  26.00&  -1.169 &  3.929&  101.9 &  86.0  \\
   5952.7180 &  26.00&  -1.439 &  3.984&   75.5 &  58.4  \\
   5956.6940 &  26.00&  -4.604 &  0.859&    ... &  63.4  \\
   5976.7770 &  26.00&  -1.242 &  3.943&   90.0 &  75.3  \\
   5984.8150 &  26.00&  -0.195 &  4.733&  106.2 &  95.3  \\
   5987.0650 &  26.00&  -0.428 &  4.796&   80.3 &   ...  \\
   6003.0120 &  26.00&  -1.119 &  3.882&  117.7 &  97.2  \\
   6007.9600 &  26.00&  -0.596 &  4.652&   70.8 &  60.0  \\
   6008.5560 &  26.00&  -0.985 &  3.884&  121.8 & 109.3  \\
   6020.1690 &  26.00&  -0.269 &  4.608&    ... & 131.0  \\
   6024.0580 &  26.00&  -0.119 &  4.549&    ... & 153.4  \\
   6027.0510 &  26.00&  -1.088 &  4.076&   80.8 &  77.5  \\
   6056.0050 &  26.00&  -0.459 &  4.733&   91.8 &  74.3  \\
   6062.8480 &  26.00&  -4.139 &  2.176&   17.9 &   ...  \\
   6078.4910 &  26.00&  -0.320 &  4.796&   97.2 &  82.4  \\
   6082.7110 &  26.00&  -3.572 &  2.223&   40.7 &  31.9  \\
   6085.2590 &  26.00&  -3.094 &  2.759&   33.3 &   ...  \\
   6093.6440 &  26.00&  -1.499 &  4.608&   30.1 &   ...  \\
   6096.6650 &  26.00&  -1.929 &  3.984&   40.8 &  34.4  \\
   6098.2450 &  26.00&  -1.879 &  4.559&   17.0 &   ...  \\
   6127.9070 &  26.00&  -1.398 &  4.143&   .... &  43.3  \\
   6151.6180 &  26.00&  -3.298 &  2.176&   65.3 &  59.4  \\
   6157.7280 &  26.00&  -1.259 &  4.076&   79.3 &  62.6  \\
   6165.3600 &  26.00&  -1.473 &  4.143&   .... &  39.1  \\
   6170.5070 &  26.00&  -0.439 &  4.796&   91.3 &  77.4  \\
   6173.3360 &  26.00&  -2.879 &  2.223&   99.9 &  89.8  \\
   6180.2040 &  26.00&  -2.585 &  2.728&   67.3 &  62.8  \\
   6187.9900 &  26.00&  -1.719 &  3.943&   43.6 &  45.2  \\
   6200.3130 &  26.00&  -2.436 &  2.609&  102.7 &  96.9  \\
   6213.4300 &  26.00&  -2.481 &  2.223&  140.4 & 131.6  \\
   6215.1440 &  26.00&  -1.319 &  4.186&   73.8 &   ...  \\
   6219.2810 &  26.00&  -2.432 &  2.198&    ... & 158.5  \\
   6220.7840 &  26.00&  -2.459 &  3.882&   18.4 &   ...  \\
   6226.7360 &  26.00&  -2.219 &  3.884&   24.3 &   ...  \\
   6229.2280 &  26.00&  -2.804 &  2.845&   56.1 &  44.5  \\
   6232.6410 &  26.00&  -1.222 &  3.654&  102.7 &  96.9  \\
   6240.6460 &  26.00&  -3.232 &  2.223&   60.1 &  49.6  \\
   6246.3190 &  26.00&  -0.732 &  3.603&  161.7 & 153.7  \\
   6256.3620 &  26.00&  -2.407 &  2.453&    ... & 134.5  \\
   6265.1340 &  26.00&  -2.549 &  2.176&    ... & 134.7  \\
   6271.2790 &  26.00&  -2.702 &  3.332&   25.1 &   ...  \\
   6290.9650 &  26.00&  -0.773 &  4.733&    ... &  54.6  \\
   6297.7930 &  26.00&  -2.739 &  2.223&    ... & 101.4  \\
   6301.5010 &  26.00&  -0.717 &  3.654&    ... & 161.6  \\
   6311.5000 &  26.00&  -3.140 &  2.832&   36.8 &  24.9  \\
   6322.6860 &  26.00&  -2.425 &  2.588&  107.2 &  96.1  \\
   6336.8240 &  26.00&  -0.855 &  3.686&  147.0 & 132.2  \\
   6344.1490 &  26.00&  -2.922 &  2.433&   69.7 &  72.1  \\
   6355.0290 &  26.00&  -2.349 &  2.845&   92.4 &  79.3  \\
   6358.6980 &  26.00&  -4.467 &  0.859&  105.4 & 100.7  \\
   6364.3660 &  26.00&  -1.429 &  4.796&   26.5 &   ...  \\
   6380.7430 &  26.00&  -1.375 &  4.186&   50.7 &  45.6  \\
   6392.5390 &  26.00&  -4.029 &  2.279&   16.7 &   ...  \\
   6408.0180 &  26.00&  -1.017 &  3.686&  135.9 &   ...  \\
   6419.9500 &  26.00&  -0.239 &  4.733&  109.5 &  88.6  \\
   6475.6240 &  26.00&  -2.941 &  2.559&   81.2 &  61.7  \\
   6481.8700 &  26.00&  -2.983 &  2.279&    ... &  82.7  \\
   6498.9390 &  26.00&  -4.698 &  0.958&   55.7 &  51.7  \\
   6518.3670 &  26.00&  -2.459 &  2.832&   74.6 &  63.2  \\
   6533.9290 &  26.00&  -1.459 &  4.559&   28.4 &  29.8  \\
   6593.8710 &  26.00&  -2.421 &  2.433&  123.2 & 112.1  \\
   6597.5610 &  26.00&  -1.069 &  4.796&   40.8 &  44.6  \\
   6609.1100 &  26.00&  -2.691 &  2.559&   78.1 &  71.9  \\
   6625.0220 &  26.00&  -5.349 &  1.011&   23.6 &   ...  \\
   6627.5450 &  26.00&  -1.679 &  4.549&   21.3 &   ...  \\
   6663.4420 &  26.00&  -2.478 &  2.424&    ... & 127.2  \\
   6703.5670 &  26.00&  -3.159 &  2.759&   30.9 &  30.5  \\
   6710.3200 &  26.00&  -4.879 &  1.485&   14.7 &   ...  \\
   6715.3830 &  26.00&  -1.639 &  4.608&   18.7 &   ...  \\
   6725.3570 &  26.00&  -2.299 &  4.103&   14.1 &   ...  \\
   6726.6660 &  26.00&  -1.132 &  4.607&   39.8 &  32.2  \\
   6733.1510 &  26.00&  -1.579 &  4.638&   24.6 &  17.4  \\
   6739.5220 &  26.00&  -4.793 &  1.557&   12.4 &   ...  \\
   6750.1530 &  26.00&  -2.620 &  2.424&  116.7 & 100.9  \\
   6752.7070 &  26.00&  -1.203 &  4.638&   29.9 &   ...  \\
   6783.7040 &  26.00&  -3.979 &  2.588&   11.4 &   ...  \\
   6786.8600 &  26.00&  -2.069 &  4.191&   24.2 &   ...  \\
   6804.0010 &  26.00&  -1.495 &  4.652&   18.4 &   ...  \\
   6806.8450 &  26.00&  -3.209 &  2.728&   37.9 &  33.1  \\
   6810.2630 &  26.00&  -0.985 &  4.607&   45.7 &  36.3  \\
   6820.3720 &  26.00&  -1.319 &  4.638&   28.5 &  32.7  \\
   6828.5910 &  26.00&  -0.919 &  4.638&   58.0 &  50.6  \\
   6837.0060 &  26.00&  -1.686 &  4.593&   19.7 &   ...  \\
   6839.8310 &  26.00&  -3.449 &  2.559&   30.3 &  23.8  \\
   6841.3390 &  26.00&  -0.749 &  4.607&   85.3 &  61.9  \\
   6842.6860 &  26.00&  -1.319 &  4.638&   33.9 &  27.5  \\
   6843.6560 &  26.00&  -0.929 &  4.549&   70.4 &  58.0  \\
   6855.1620 &  26.00&  -0.741 &  4.559&    ... &  75.4  \\
   6858.1500 &  26.00&  -0.929 &  4.608&   58.3 &  36.1  \\
   7024.6430 &  26.00&  -1.079 &  4.559&    ... &  38.0  \\
   7038.2230 &  26.00&  -1.299 &  4.218&    ... &  57.4  \\
   7068.4100 &  26.00&  -1.379 &  4.076&   71.6 &  54.8  \\
   7090.3840 &  26.00&  -1.209 &  4.231&   69.2 &  62.0  \\
   7127.5680 &  26.00&  -1.045 &  4.988&   24.8 &   ...  \\
   7130.9220 &  26.00&  -0.789 &  4.218&  112.3 &  98.8  \\
   7132.9860 &  26.00&  -1.627 &  4.076&    ... &  32.5  \\
   7151.5000 &  26.00&  -3.729 &  2.484&   28.0 &   ...  \\
   7189.1450 &  26.00&  -2.770 &  3.071&   49.0 &  31.1  \\
   7219.6850 &  26.00&  -1.689 &  4.076&   34.0 &  33.5  \\
   7284.8350 &  26.00&  -1.749 &  4.143&   29.8 &  32.6  \\
   7306.5620 &  26.00&  -1.739 &  4.178&   40.1 &  36.2  \\
   7351.5120 &  26.00&  -0.636 &  4.956&   52.7 &   ...  \\
   7411.1530 &  26.00&  -0.298 &  4.283&  138.7 & 135.0  \\
   7440.9120 &  26.00&  -0.572 &  4.913&    ... &  44.5  \\
   7443.0220 &  26.00&  -1.819 &  4.186&   29.2 &   ...  \\
   7461.5210 &  26.00&  -3.579 &  2.559&   22.8 &   ...  \\
   7491.6470 &  26.00&  -0.898 &  4.301&   75.3 &  73.3  \\
   7568.8990 &  26.00&  -0.772 &  4.283&    ... &  76.7  \\
   7583.7880 &  26.00&  -1.884 &  3.018&  107.0 & 102.4  \\
   7586.0180 &  26.00&  -0.457 &  4.313&    ... & 151.2  \\
   7710.3650 &  26.00&  -1.112 &  4.220&    ... &  67.7  \\
   7748.2690 &  26.00&  -1.750 &  2.949&  150.6 & 141.3  \\
   7751.1090 &  26.00&  -0.752 &  4.991&   45.4 &  33.5  \\
   7780.5570 &  26.00&   0.029 &  4.473&  149.8 & 164.3  \\
   7807.9090 &  26.00&  -0.540 &  4.991&   52.6 &   ...  \\
   7832.1950 &  26.00&   0.111 &  4.435&    ... & 171.1  \\
   7855.3990 &  26.00&  -1.019 &  5.064&   26.1 &   ...  \\
   5120.3520 &  26.01&  -4.255 &  2.828&    ... &  90.7  \\
   5132.6690 &  26.01&  -3.979 &  2.807&    ... &  98.3  \\
   5161.1840 &  26.01&  -4.572 &  2.856&    ... &  49.0  \\
   5256.9370 &  26.01&  -4.181 &  2.891&    ... &  94.3  \\
   5414.0730 &  26.01&  -3.539 &  3.221&    ... & 128.4  \\
   5425.2570 &  26.01&  -3.159 &  3.199&    ... & 156.7  \\
   5627.4970 &  26.01&  -4.129 &  3.387&    ... &  57.4  \\
   5991.3760 &  26.01&  -3.539 &  3.153&    ... & 132.0  \\
   6084.1110 &  26.01&  -3.779 &  3.199&    ... &  96.5  \\
   6113.3220 &  26.01&  -4.109 &  3.221&   80.9 &  73.5  \\
   6149.2580 &  26.01&  -2.719 &  3.889&    ... & 143.1  \\
   6233.5340 &  26.01&  -2.831 &  5.484&   11.5 &   ...  \\
   6369.4620 &  26.01&  -4.159 &  2.891&    ... & 100.9  \\
   6383.7220 &  26.01&  -2.069 &  5.553&   47.3 &  31.6  \\
   6416.9190 &  26.01&  -2.649 &  3.892&    ... & 145.9  \\
   6432.6800 &  26.01&  -3.519 &  2.891&    ... & 155.8  \\
   6442.9550 &  26.01&  -2.670 &  5.549&    ... &  17.7  \\
   6446.4100 &  26.01&  -1.959 &  6.223&   23.0 &   ...  \\
   7222.3940 &  26.01&  -3.359 &  3.889&    ... &  84.8  \\
   7449.3350 &  26.01&  -3.089 &  3.889&  102.2 &        \\
   7479.6930 &  26.01&  -3.679 &  3.892&   55.7 &  44.7  \\
   7515.8310 &  26.01&  -3.459 &  3.903&   76.9 &  71.6  \\
   7711.7230 &  26.01&  -2.499 &  3.903&  161.8 &   ...  \\
   5483.3530 &  27.00&  -1.489 &  1.710&    ... &  59.4  \\
   5647.2340 &  27.00&  -1.559 &  2.280&   23.0 &   ...  \\
   6188.9960 &  27.00&  -2.449 &  1.710&   19.4 &   ...  \\
   6771.0340 &  27.00&  -1.969 &  1.883&   15.4 &  16.5  \\
   6814.9440 &  27.00&  -1.899 &  1.956&   22.5 &   ...  \\
   7084.9830 &  27.00&  -1.114 &  1.883&   70.4 &  54.9  \\
   5000.3430 &  28.00&  -0.429 &  3.635&    ... & 102.7  \\
   5003.7410 &  28.00&  -2.799 &  1.676&    ... &  50.3  \\
   5010.9380 &  28.00&  -0.869 &  3.635&    ... &  51.6  \\
   5032.7270 &  28.00&  -1.269 &  3.898&    ... &  15.5  \\
   5035.3570 &  28.00&   0.290 &  3.635&    ... & 162.8  \\
   5048.8470 &  28.00&  -0.379 &  3.847&    ... &  77.4  \\
   5080.5280 &  28.00&   0.330 &  3.655&    ... & 171.7  \\
   5081.1100 &  28.00&   0.300 &  3.847&    ... & 142.3  \\
   5082.3440 &  28.00&  -0.539 &  3.658&    ... &  77.1  \\
   5084.0960 &  28.00&   0.030 &  3.679&    ... & 133.1  \\
   5099.9300 &  28.00&  -0.099 &  3.679&    ... & 121.5  \\
   5115.3920 &  28.00&  -0.109 &  3.834&    ... & 108.1  \\
   5155.1260 &  28.00&  -0.649 &  3.898&    ... &  46.5  \\
   5155.7640 &  28.00&   0.074 &  3.898&    ... &  91.4  \\
   5176.5600 &  28.00&  -0.439 &  3.898&    ... &  57.7  \\
   5435.8580 &  28.00&  -2.589 &  1.986&    ... &  53.5  \\
   5578.7180 &  28.00&  -2.639 &  1.676&   69.6 &  63.3  \\
   5587.8580 &  28.00&  -2.139 &  1.935&    ... &  70.0  \\
   5593.7350 &  28.00&  -0.839 &  3.898&   49.7 &  35.5  \\
   5663.9850 &  28.00&  -0.429 &  4.538&    ... &  25.0  \\
   5694.9830 &  28.00&  -0.609 &  4.089&    ... &  37.4  \\
   5754.6560 &  28.00&  -2.329 &  1.935&  106.2 &   ...  \\
   5760.8300 &  28.00&  -0.799 &  4.105&   34.0 &   ...  \\
   5805.2170 &  28.00&  -0.639 &  4.167&   44.5 &  31.4  \\
   5831.5950 &  28.00&  -0.944 &  4.167&   25.3 &  21.3  \\
   5846.9930 &  28.00&  -3.209 &  1.676&   26.4 &  21.3  \\
   5996.7300 &  28.00&  -1.059 &  4.236&   16.2 &   ...  \\
   6086.2810 &  28.00&  -0.529 &  4.266&   28.0 &  29.1  \\
   6108.1160 &  28.00&  -2.449 &  1.676&   88.7 &  90.4  \\
   6111.0700 &  28.00&  -0.869 &  4.088&   35.6 &  22.3  \\
   6128.9730 &  28.00&  -3.329 &  1.676&   28.2 &   ...  \\
   6133.9630 &  28.00&  -1.829 &  4.088&    7.1 &   ...  \\
   6175.3660 &  28.00&  -0.529 &  4.089&   67.2 &  48.6  \\
   6176.8070 &  28.00&  -0.259 &  4.088&   78.4 &  64.3  \\
   6204.6000 &  28.00&  -1.099 &  4.088&   21.6 &   ...  \\
   6223.9810 &  28.00&  -0.909 &  4.105&   25.4 &  21.7  \\
   6259.5950 &  28.00&  -1.236 &  4.089&   17.4 &   ...  \\
   6322.1660 &  28.00&  -1.169 &  4.154&   25.8 &   ...  \\
   6327.5980 &  28.00&  -3.149 &  1.676&   42.4 &  37.7  \\
   6360.8110 &  28.00&  -1.026 &  4.167&   15.5 &   ...  \\
   6366.4800 &  28.00&  -0.873 &  4.167&   17.7 &   ...  \\
   6378.2470 &  28.00&  -0.829 &  4.154&   33.3 &  28.1  \\
   6532.8730 &  28.00&  -3.389 &  1.935&   23.9 &   ...  \\
   6586.3100 &  28.00&  -2.809 &  1.951&   53.0 &  37.7  \\
   6598.5980 &  28.00&  -0.979 &  4.236&   30.4 &   ...  \\
   6635.1220 &  28.00&  -0.819 &  4.419&   16.8 &   ...  \\
   6767.7720 &  28.00&  -2.169 &  1.826&  137.5 & 120.5  \\
   6772.3150 &  28.00&  -0.979 &  3.658&   58.7 &  46.3  \\
   6914.5590 &  28.00&  -2.269 &  1.951&    ... &  86.3  \\
   7110.8960 &  28.00&  -2.979 &  1.935&   35.9 &   ...  \\
   7122.1970 &  28.00&   0.040 &  3.542&  159.0 & 153.9  \\
   7181.9690 &  28.00&  -0.739 &  3.743&   79.3 &  74.4  \\
   7385.2370 &  28.00&  -1.969 &  2.740&   38.6 &   ...  \\
   7422.2750 &  28.00&  -0.139 &  3.635&  134.7 & 139.5  \\
   7522.7590 &  28.00&  -0.464 &  3.658&   95.6 &  90.7  \\
   7525.1110 &  28.00&  -0.432 &  3.635&   99.8 &  82.1  \\
   7555.5970 &  28.00&   0.054 &  3.847&  129.3 & 120.0  \\
   7574.0420 &  28.00&  -0.448 &  3.833&    ... &  74.8  \\
   7797.5800 &  28.00&  -0.184 &  3.898&  102.9 &  97.1  \\
   4722.1530 &  30.00&  -0.337 &  4.030&    ... & 143.8  \\
   4810.5280 &  30.00&  -0.136 &  4.078&    ... & 163.0  \\
   5087.4160 &  39.01&  -0.169 &  1.084&    ... &  75.9  \\
   5200.4060 &  39.01&  -0.569 &  0.992&    ... &  54.1  \\
   5509.8950 &  39.01&  -0.947 &  0.992&    ... &  39.7  \\
   6613.7320 &  39.01&  -0.847 &  1.748&   27.6 &   ...  \\
   6795.4140 &  39.01&  -1.029 &  1.738&   14.9 &   ...  \\
   6407.2160 &  40.01&  -2.699 &  0.154&   55.7 &   ...  \\
   6262.2900 &  57.01&  -1.219 &  0.403&   19.6 &   ...  \\
   6320.3760 &  57.01&  -1.609 &  0.173&   19.8 &   ...  \\
   6390.4800 &  57.01&  -1.409 &  0.321&   17.8 &   ...  \\
   5092.7900 &  60.01&  -0.609 &  0.380&    ... &  32.3  \\
   5130.5900 &  60.01&   0.450 &  1.304&    ... &  32.5  \\
   5740.8600 &  60.01&  -0.529 &  1.160&   11.5 &   ...  \\
   6645.0940 &  63.01&  -0.161 &  1.380&   66.5 &   ...  \\
\end{longtable}  

\end{appendix}

\end{document}